\documentclass[final,5p,times,twocolumn, number, sort&compress]{elsarticle}

\usepackage{setup}
\usepackage{amssymb}
\usepackage{lipsum}
\usepackage{graphicx}
\usepackage{bm}
\usepackage{subcaption}
\usepackage{mathtools}
\usepackage{comment}
\usepackage[style=plain]{floatrow}
\usepackage{xspace}
\usepackage[colorlinks=true]{hyperref}
\usepackage[utf8]{inputenc}    
\usepackage[T1]{fontenc}

\makeatletter

\newcommand{\preprint}[1]{\gdef\@preprint{#1}}
\def\@preprint{}

\def\ps@pprintTitle{%
	\let\@oddfoot\@empty
	\let\@evenfoot\@empty
	\def\@oddhead{%
		\vbox to 0pt{%
			\vskip -1cm 
			\hfill
			\parbox[t]{1\textwidth}{\raggedleft\@preprint}%
			\vss
		}%
	}%
	\let\@evenhead\@oddhead
}

\makeatother

\allowdisplaybreaks
\journal{Physics Letters B}
\def\lapprox{\lower .7ex\hbox{$\;\stackrel{\textstyle <}{\sim}\;$}}
\def\gapprox{\lower .7ex\hbox{$\;\stackrel{\textstyle >}{\sim}\;$}}
\definecolor{lightgray}{HTML}{A6A39A}
\definecolor{darkgray}{HTML}{504E48}
\definecolor{silver}{HTML}{E0DFDE}
\definecolor{brown}{HTML}{5F4541}
\definecolor{beige}{HTML}{DCCCAC}
\definecolor{green}{HTML}{345F53}
\definecolor{yellow}{HTML}{F6B65A}
\definecolor{blue}{HTML}{568BCF}
\definecolor{red}{HTML}{AE1932}
\definecolor{orange}{HTML}{D16F15}
\definecolor{darkpurple}{HTML}{39FF14}

\newcommand{\expo}[1]{\ensuremath{\mathrm{e}^{#1}}}
\newcommand{\calF}{\ensuremath{\mathcal{F}}}

\DeclareRobustCommand{\nnlojet}{\texorpdfstring{NNLO\protect\scalebox{0.8}{JET}}{NNLOJET}\xspace}
\DeclareRobustCommand{\NNLOJET}{\nnlojet}

\DeclareRobustCommand{\ARES}{\texorpdfstring{A\protect\scalebox{0.8}{RES}}{ARES}\xspace}
\DeclareRobustCommand{\CAESAR}{\texorpdfstring{C\protect\scalebox{0.8}{AESAR}}{CAESAR}\xspace}

\begin{document}

\begin{frontmatter}
	
\preprint{IPPP/25/85, ZU-TH 82/25, MCNET-25-27}

\title{NNLO+NNLL Predictions for Heavy-Jet Mass and $C$-parameter in Higgs Decays to Quarks and Gluons}

\author[a]{Elliot Fox}
\author[b,c]{Aude Gehrmann-De Ridder}
\author[c]{Thomas Gehrmann}
\author[a,d]{Nigel Glover}
\author[a]{Matteo Marcoli}
\author[e]{Christian T. Preuss}

\affiliation[a]{organization={Institute for Particle Physics Phenomenology, Department of Physics, Durham University},
city={Durham},
postcode={DH1 3LE}, 
country={UK}}

\affiliation[b]{organization={Institute for Theoretical Physics, ETH},
city={Z\"urich},
postcode={8093}, 
country={Switzerland}}
	
\affiliation[c]{organization={Physik-Institut, Universit\"{a}t Z\"{u}rich},
city={Z\"{u}rich},
postcode={8057}, 
country={Switzerland}}
	
\affiliation[d]{organization={CERN},
city={Geneva},
postcode={1211}, 
country={Switzerland}}
            
\affiliation[e]{organization={Institut f{\"u}r Theoretische Physik, Georg-August-Universit{\"a}t G{\"o}ttingen},
city={G{\"o}ttingen},
postcode={ 37077}, 
country={Germany}}

\begin{abstract}
  We consider the resummation of large logarithmic corrections arising in the two-particle limit at next-to-next-to-leading logarithmic (NNLL) accuracy for the heavy-jet mass and $C$-parameter distributions in the decay of a Higgs boson to quarks and gluons: $H\to b\bar{b}$, $H\to c\bar{c}$, and $H\to gg$.
  We demonstrate how the matched NNLO+NNLL results clarify the relative contributions of key hadronic Higgs-decay channels ($H\to b\bar{b}$, $H\to c\bar{c}$, $H\to gg$) yielding reduced uncertainties for both event-shape observables -- especially for heavy-jet mass -- while revealing substantial effects that shift the $C$-parameter peak in gluonic decays.
\end{abstract}

\end{frontmatter}

\section{Introduction}

Precise determinations of Higgs-boson properties will become possible in measurements at future lepton colliders~\cite{FCC:2018byv,FCC:2018evy,CEPCStudyGroup:2018ghi,ILC:2013jhg}, which will be operated as ``Higgs factories'' to produce large numbers of Higgs bosons.
This has led to an increased interest in precise theoretical predictions of event-shape observables in hadronic Higgs decays \cite{Alioli:2020fzf,Coloretti:2022jcl,Gao:2016jcm,Gao:2019mlt,Gao:2020vyx,Gehrmann-DeRidder:2023uld,Gehrmann-DeRidder:2024avt,Fox:2025cuz,Fox:2025qmp,Fox:2025txz,Ju:2023dfa,Knobbe:2023njd,Luo:2019nig,Mo:2017gzp,vanBeekveld:2024wws,Yan:2023xsd,Zhu:2023oka}.
In \cite{Fox:2025txz}, we have recently obtained the thrust distribution in hadronic Higgs decays at NNLO+NNLL accuracy.
Here, we consider two related observables, the heavy-jet mass and the $C$-parameter, and match the numerical NNLO calculation from \cite{Fox:2025qmp} to the analytical NNLL resummation in the two-particle limit obtained analytically with the \ARES formalism \cite{Banfi:2014sua,Banfi:2018mcq}.

The heavy-jet mass event shape is defined using the two hemispheres $\mathcal{H}_\mathrm{L}$ and $\mathcal{H}_\mathrm{R}$, identified by the plane orthogonal to the thrust axis~\cite{Clavelli:1981yh},
\begin{equation}
  \rho_\mathrm{H} \equiv \frac{M_\mathrm{H}^2}{s} = \max_{i\in\{\mathrm{L},\mathrm{R}\}}\left(\frac{M_i^2}{s}\right) \,,
\end{equation}
where the scaled invariant hemisphere masses $M_{\mathrm{L}/\mathrm{R}}$ are given by
\begin{equation}
  \frac{M_{\mathrm{L}/\mathrm{R}}^2}{s} = \frac{1}{s}\left(\sum\limits_{j \in \mathcal{H}_{\mathrm{L}/\mathrm{R}}} p_j\right)^2 \,,
\end{equation}
with $p_j$ denoting the four-momenta of the final-state particles.
At LO and LL, the heavy-jet mass and thrust event shapes are identical for massless particles, but they differ starting from NLO and NLL.

The $C$-parameter is defined using the three eigenvalues $\lambda_1$, $\lambda_2$, and $\lambda_3$ of the linearised momentum tensor \cite{Parisi:1978eg,Donoghue:1979vi},
\begin{equation}
  \Theta^{\alpha\beta} = \frac{1}{\sum\limits_j\mods{\vec{p}_j}}\sum\limits_i\frac{p_i^\alpha p_i^\beta}{\mods{\vec{p}_i}} \, ,\;\; \text{where } \alpha,\beta \in \{1,2,3\} \,,
\end{equation}
as
\begin{equation}
  C = 3(\lambda_1\lambda_2 + \lambda_2\lambda_3 + \lambda_3\lambda_1) \,.
\end{equation}
The $\lambda_i$ satisfy $0\leq \lambda_{1,2,3}\leq 1$ and $\sum_i\lambda_i=1$, therefore \mbox{$0\leq C\leq 1$}.
For massless particles, the resummation procedure for the \mbox{$C$-parameter} is identical to thrust up to NLL, upon replacing \mbox{$C \to 6\tau$}, but differs from the thrust result at NNLL due to wide-angle soft radiation.
At fixed order, the $C$-parameter and thrust event shapes differ starting at LO.

Due to their importance in LEP measurements, see e.g.~\cite{JADE:1999zar,ALEPH:2003obs,DELPHI:2003yqh,L3:2004cdh}, both the heavy-jet mass and the $C$-parameter have extensively been studied in theory and experiments in electron-positron annihilation.
On the theory side, the resummation of the $C$-parameter has been achieved at NLL in \cite{Catani:1998sf}, extended to NNLL \cite{Banfi:2014sua} and performed up to N$^3$LL in \cite{Hoang:2014wka}.
Subleading-power corrections to the $C$-parameter distribution have been studied in \cite{Agarwal:2023fdk,Buonocore:2023mne,Hoang:2014wka} and non-perturbative aspects have been considered in \cite{Gardi:2003iv,Luisoni:2020efy,Caola:2022vea,Caola:2021kzt}.
Precise determinations of the strong coupling from fitting the $C$-parameter distribution to LEP data have been obtained in \cite{Hoang:2015hka,Luisoni:2020efy,Nason:2023asn,Nason:2025qbx}.
The $C$-parameter contains an integrable singularity at the kinematical three-particle phase-space boundary, leading to large Sudakov-shoulder logarithms \cite{Catani:1997xc}.

For the heavy-jet mass, the first NLL resummation in electron-positron annihilation has been obtained in \cite{Catani:1992ua} and later extended to NNLL \cite{Chien:2010kc,Banfi:2014sua} and N$^3$LL, including subleading-power corrections, in \cite{Hoang:2025uaa}.
The heavy-jet mass observable is special, as it entails a two-sided Sudakov shoulder, which gives rise to large logarithmic contributions that can be resummed up to NNLL \cite{Bhattacharya:2022dtm,Bhattacharya:2023qet}.
Fixed-order predictions for both event-shape observables have first been obtained at NLO in \cite{Ellis:1980wv} and at NNLO in \cite{Gehrmann-DeRidder:2007vsv,Weinzierl:2009nz,DelDuca:2016ily}, and non-perturbative corrections were analyzed in \cite{Dokshitzer:1997ew,Gardi:2002bg,Caola:2022vea,Hoang:2025uaa}.
Heavy-jet mass data from LEP and earlier $e^+e^-$ colliders were used in recent extractions of the strong coupling constant \cite{Benitez:2025vsp,Nason:2023asn,Nason:2025qbx}.

In hadronic Higgs decays to quarks and gluons, NLO predictions for both the heavy-jet mass and the $C$-parameter have been obtained for the first time in \cite{Coloretti:2022jcl}, at NNLO in \cite{Fox:2025qmp}, and at NLO+NLL$^\prime$ accuracy in \cite{Gehrmann-DeRidder:2024avt}.

This letter is structured as follows.
In Section~\ref{sec:calculation}, we describe the calculation of the event-shape observables at NNLO and NNLL, before showing the matched distributions in $H\to q\bar{q}$ and $H\to gg$ decays in Section~\ref{sec:results}.
We conclude and give an outlook on future developments in Section~\ref{sec:conclusions}.

\section{Calculation}
\label{sec:calculation}

\begin{figure}[t]
  \centering
  \includegraphics[width=\textwidth]{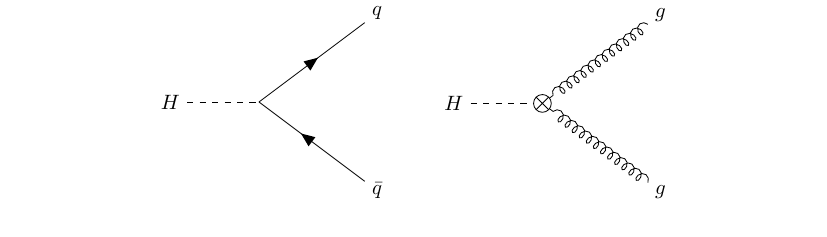}
  \caption{Hadronic Higgs decay categories: $H\to q\bar{q}$ with a Yukawa coupling  (left) and $H\to gg$ via an effective coupling (right).}
  \label{fig:diagH2jLO}
\end{figure}

The leading-order inclusive widths for a Higgs boson decaying via a Yukawa interaction to a quark-antiquark pair or to gluons via an effective vertex~\cite{Wilczek:1977zn} read
\begin{equation}
  \GammaHqq^{(0)} = \frac{y_q^2(\muR)m_\PH\NC}{8\uppi} \,, \quad \GammaHgg^{(0)} = \frac{\lambda_0^2(\muR)m_\PH^3(\NC^2-1)}{64\uppi} \, ,
  \label{eq:ratesLO}
\end{equation}
where the Yukawa coupling $y_q$ and the effective coupling $\lambda_0$ are given in terms of the Fermi constant $\GF$  as
\begin{equation}\label{eq:couplings}
  y_b^2(\muR) = m_b^2(\muR)\sqrt{2}\GF\,, \quad \lambda_0^2(\muR) = \frac{\alphas^2(\muR)\sqrt{2}\GF}{9\uppi^2}\,.
\end{equation}
Higher-order QCD corrections to the inclusive widths are given by
\begin{align}
  \Gamma^{(k)}_{H\to X} &= \Gamma^{(0)}_{H\to X}\, \left(1+\sum\limits_{n=1}^k \as^n C_{X}^{(n)}\right)\,\quad\text{with}\,X=q\bar{q},gg\,
  \label{eq:ratesNkLO}
\end{align}
where $\as=\alphas/(2\uppi)$.
The perturbative coefficients $C_{X}^{(n)}$ are computed up to fourth order in~\cite{Herzog:2017dtz}.

For a generic event shape observable $y$, we consider the cumulant of the differential decay rate
\begin{align}
  \Sigma_{H\to X}(y) &= \frac{1}{\Gamma_{H\to X}}\int\limits^{y}_0\, \frac{\rd \Gamma_{H\to X}}{\rd y'}\, \rd y' \,.
  \label{eq:cumulant1}
\end{align}
This quantity can be computed in perturbation theory as
\begin{equation}
  \Sigma(y) = 1 + \as\mathcal{A}(y) + \as^2\mathcal{B}(y) + \as^3\mathcal{C}(y) + \order{\as^4}\,,
  \label{eq:cumulantFO}
\end{equation}
where the coefficients $\mathcal{A}$, $\mathcal{B}$ and $\mathcal{C}$ can be obtained by a fixed-order calculation, respectively at LO, NLO and NNLO, for the decay of a Higgs boson to three partons, in combination with the knowledge of the inclusive decay width at the same order in $\alphas$. The calculation requires matrix elements for the decay of a Higgs boson to up to five partons at 
tree-level~\cite{DelDuca:2004wt,Anastasiou:2011qx,DelDuca:2015zqa,Mondini:2019vub,Mondini:2019gid}, up to four partons at one loop~\cite{Dixon:2009uk,Badger:2009hw,Badger:2009vh,Anastasiou:2011qx,DelDuca:2015zqa,Mondini:2019vub,Mondini:2019gid} and three partons at two loops~\cite{Ahmed:2014pka,Mondini:2019vub,Chen:2014gva}. Moreover, to suitably combine these in physical predictions, dedicated techniques must be employed to treat infrared singularities which appear in intermediate stages of higher-order calculations. We rely on the antenna subtraction method~\cite{Gehrmann-DeRidder:2005btv,Currie:2013vh} to compute the coefficients in~\eqref{eq:cumulantFO} up to NNLO accuracy. In particular, we make use of generalised antenna functions~\cite{Fox:2024bfp} constructed with the algorithm presented in~\cite{Braun-White:2023sgd,Braun-White:2023zwd}. All these ingredients are implemented within the \NNLOJET~\cite{NNLOJET:2025rno} parton-level event generator which has been suitably extended to include hadronic Higgs decay processes~\cite{Fox:2025cuz,Fox:2025qmp}.

For intermediate and large values of the event shape $y$, the results obtained using the procedure detailed above yield reliable theoretical predictions.
However, it is well-known that as $y\to 0$, corresponding to the two-particle limit, the emergence of large logarithmic corrections of the form $\alphas^n \log(y)^m$ at each order in perturbation theory leads to unphysical results.
To restore predictive power, such contributions need to be resummed to all orders in the strong coupling.
We define $L\equiv-\log(y)$ and $\lambda\equiv\alphas\beta_0L$, where $\beta_0$ is the first-order coefficient of the QCD $\beta$-function.
As both the heavy-jet mass and $C$-parameter event shapes factorise, the cumulant distribution can be written in the $y\to 0$ limit as
\begin{align}
  \Sigma_{H\to X}(y) = \left(1+\sum\limits_n \as^n c_{n}^{(H\to X)}\right)\expo{-R_X(\lambda)}\calF_X(R_X^\prime(\lambda))\,,
  \label{eq:cumulant}
\end{align}
where $R_X$ is the Sudakov radiator which describes single-emission contributions, while the $\calF_X$ function accounts for multiple emissions.
Both quantities only depend on the particle type of the hard emitters and on the observable $y$, and they encode the all-orders behaviour of $\Sigma(y)$.
The perturbative coefficients $c_{n}^{(H\to X)}$, on the other hand, are process-dependent, contribute only non-logarithmic terms, and induce an overall normalisation.
We compute these ingredients up to NNLL accuracy relying on the \ARES formalism~\cite{Banfi:2014sua,Banfi:2018mcq}, following the steps detailed for the thrust event shape in~\cite{Fox:2025txz}.

General expressions for the NNLL Sudakov radiator for quark and gluon legs are given in \cite{Banfi:2018mcq}.
As both event shapes are additive, we can utilise known results for the multiple-emission function $\calF_X$ of additive observables at NLL \cite{Catani:1992ua} and its NNLL corrections for quark radiators \cite{Banfi:2014sua}.
For gluon radiators, we have derived the relevant NNLL corrections following the \ARES formalism detailed in \cite{Banfi:2014sua,Banfi:2018mcq} and cross checked against \cite{Arpino:2019ozn}, where applicable.
Up to NNLL, the heavy-jet mass distribution can then be obtained as the product of the two hemisphere mass distributions \cite{Chien:2010kc}, suitably expanded to include only pure NNLL contributions.
For the $C$-parameter distribution at NNLL, the thrust result derived in \cite{Fox:2025txz} can be extended to account for the overall normalisation difference and additional soft wide-angle contributions.
As in the thrust case \cite{Fox:2025txz}, we eliminate non-logarithmic contributions at $\mathcal{O}(\alphas)$ from $\calF_X$ in favour of including the $c_n^{(H\to X)}$ coefficients in~\eqref{eq:cumulant}.

\begin{figure*}[t]
  \centering
  \hspace{-0.5cm}\includegraphics[width=0.3\textwidth]{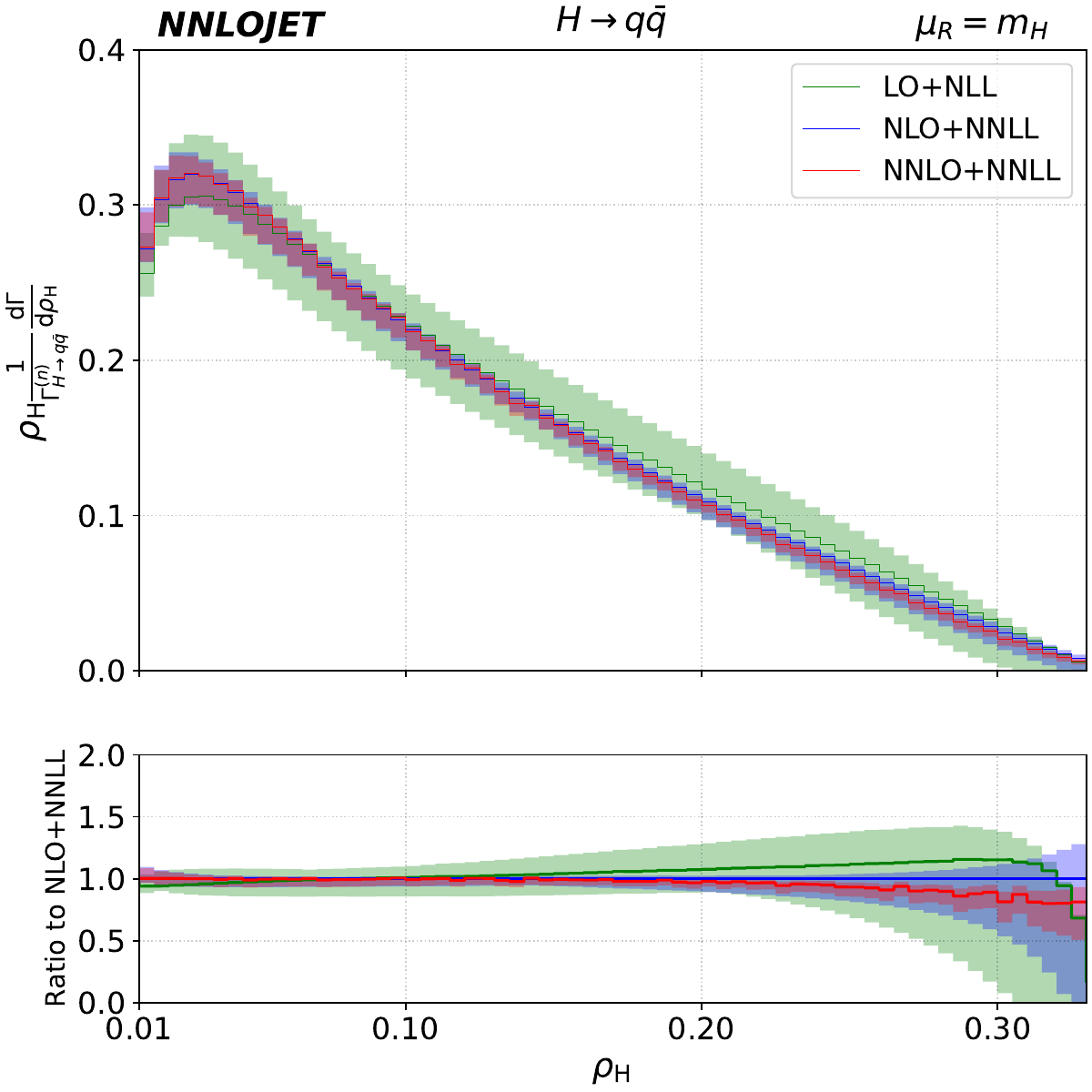}
  \hspace{0.3cm}
  \includegraphics[width=0.3\textwidth]{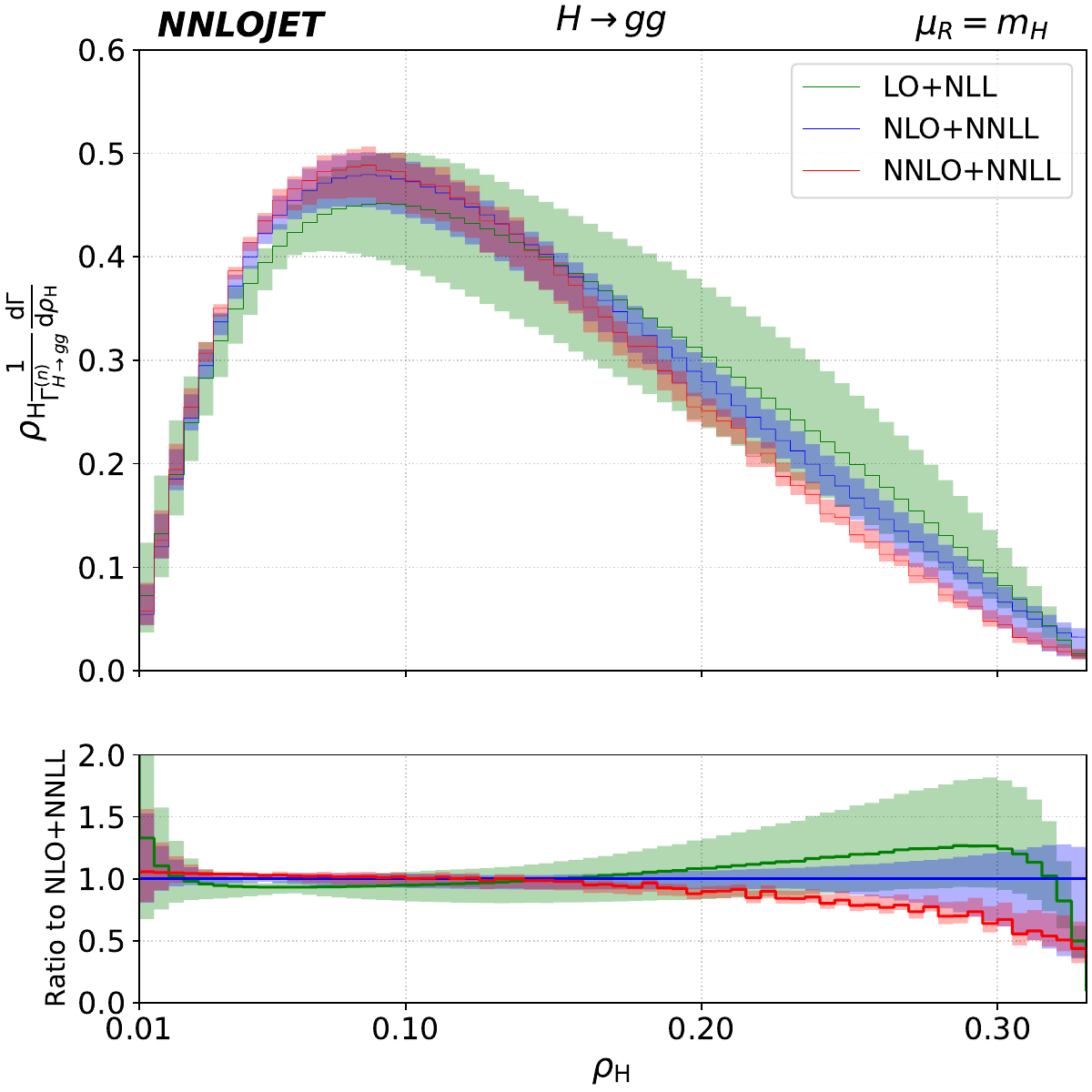}
  \hspace{0.3cm}
  \includegraphics[width=0.3\textwidth]{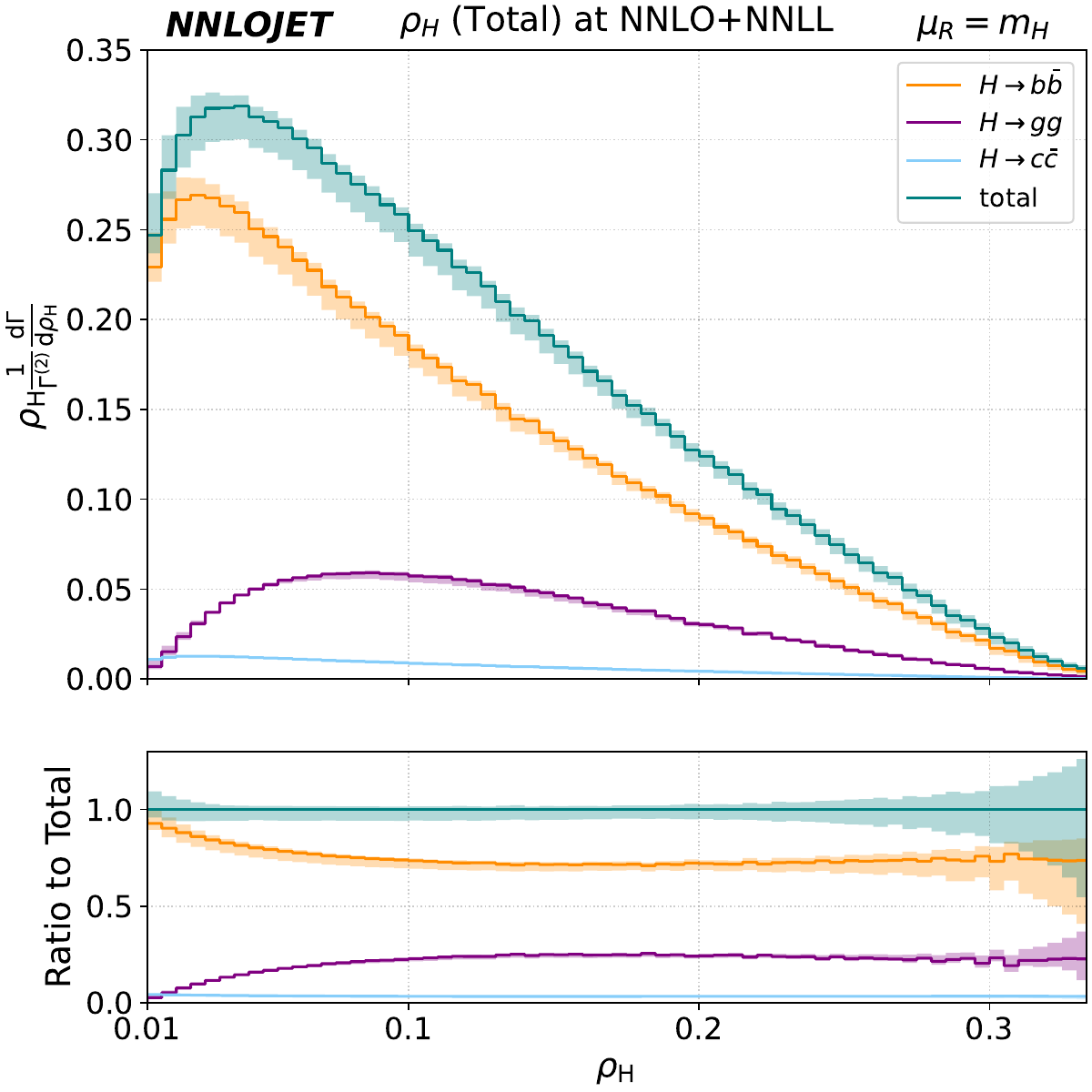}\\
  \hspace{-0.5cm}\includegraphics[width=0.3\textwidth]{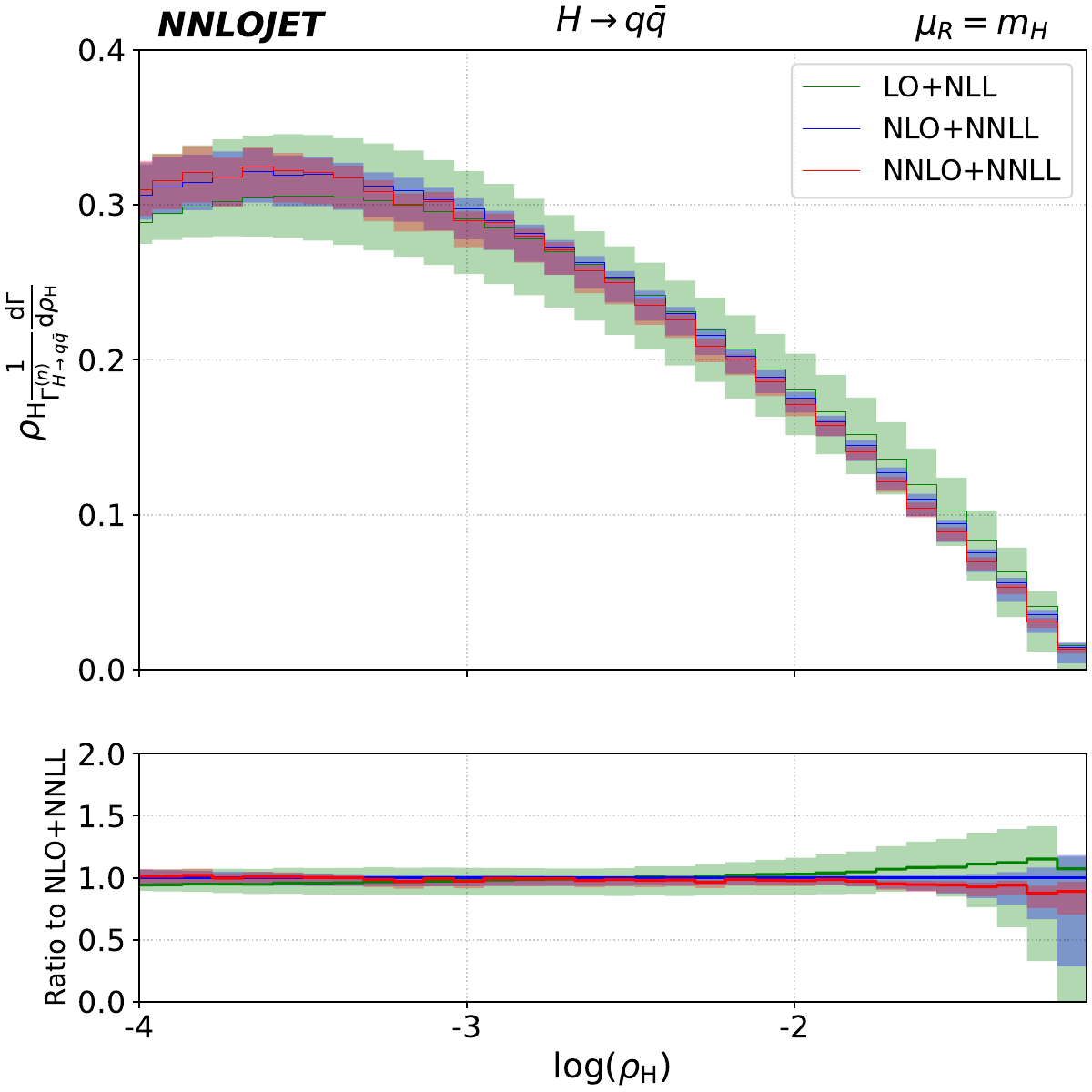}
  \hspace{0.3cm}
  \includegraphics[width=0.3\textwidth]{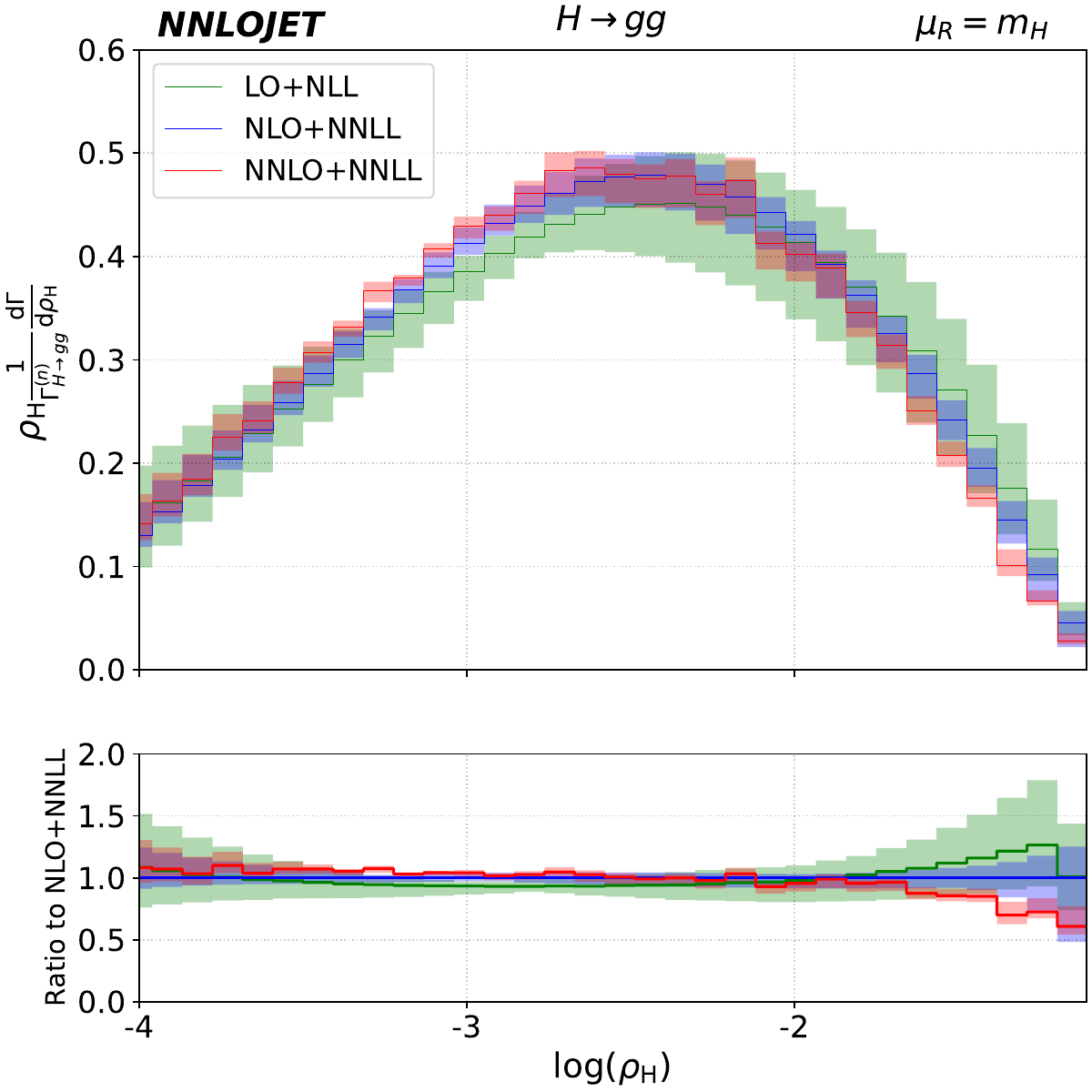}
  \hspace{0.3cm}
  \includegraphics[width=0.3\textwidth]{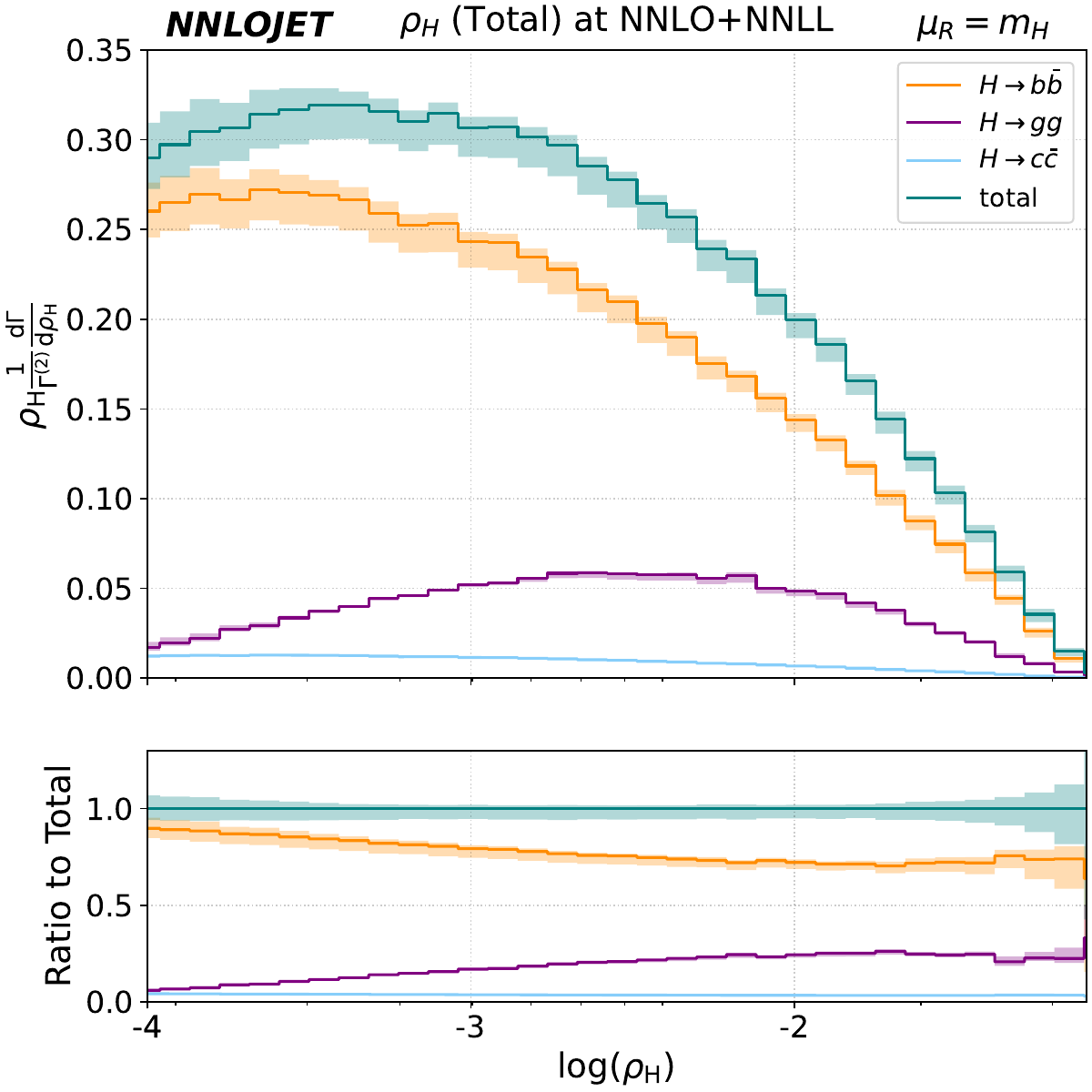}
  \caption{Heavy-jet mass distribution results for fixed order predictions matched to resummation in the logR scheme for the $H\to q\bar{q}$ channel (left column), $H\to gg$ channel (middle column) and the sum over all channels (right column) for both linear-binning (top row) and logarithmic-binning (bottom row). For the frames in the two leftmost columns, the curves represent the LO calculation matched to NLL resummation (green), NLO matched to NNLL (blue), and NNLO matched to NNLL (red), and the ratio to NLO+NNLL is shown in the lower frame. For the frames in the rightmost column, the curves represent the sum over all decay channels at NNLO matched to NNLL (teal), the $H\to b\bar{b}$ channel (orange), the $H\to gg$ channel (purple) and the $H \to c\bar{c}$ channel (light blue), and the ratio to the sum is shown in the lower frame. }
  \label{fig:HJM}
\end{figure*}

Up to $\alphas^3$, the fixed-order expansions of the LL, NLL, and NNLL cumulant are given by
\begin{align}
  \Sigma_\text{LL}(y) &= \exp\left\{G_{12}a_sL^2 + G_{23}a_s^2L^3 + G_{34}a_s^3L^4 + \ldots\right\}\,, \\
  \Sigma_\text{NLL}(y) &= \exp\left\{G_{11}a_sL + G_{22}a_s^2L^2 + G_{33}a_s^3L^3 + \ldots\right\} \,, \\
  \Sigma_\text{NNLL}(y) &= \exp\left\{G_{21}a_s^2L + G_{32}a_s^3L^2 + \ldots\right\} \,.
\end{align}
The $G_{ij}$ and $c_1^{(H\to X)}$ coefficients for the event shape observables considered in this letter are given in \ref{app:Gij}.
We have validated the correctness of the expansion coefficients analytically up to NNLL against a previous calculation in the $H\to q\bar{q}$ decay mode~\cite{Chien:2010kc,Hoang:2014wka} and up to NLL against the \CAESAR result for $H\to gg$ decays~\cite{Banfi:2004yd}.

To obtain predictions valid across all values of $y$, the resummed results must be suitably combined with the fixed-order calculations.
This is achieved via a matching procedure which adds the two descriptions, while removing contributions which are doubly counted.
We perform the matching of the NNLL resummation to the NNLO predictions using the logR scheme~\cite{Catani:1992ua,Jones:2003yv},
\begin{equation}
	\label{eq:matching}
  \begin{split}
    \log\left(\Sigma(y)\right) &= -R_\text{NNLL}(y) + \log\left(\calF_\text{NNLL}(y)\right) \\
    &\hspace{-1.3cm}+a_s\left(\mathcal{A}(y) - G_{11}L - G_{12}L^2\right) \\
    &\hspace{-1.3cm}+a_s^2\left(\mathcal{B}(y) - \frac{1}{2}\mathcal{A}(y)^2 - G_{21}L - G_{22}L^2 - G_{23}L^3\right) \\
    &\hspace{-1.3cm}+ a_s^3\left(\mathcal{C}(y) - \mathcal{A}(y)\mathcal{B}(y) + \frac{1}{3}\mathcal{A}(y)^3 - G_{32}L^2 - G_{33}L^3 - G_{34}L^4\right)\,. \\
  \end{split}
\end{equation}
Up to NLO, the fixed-order expansion of the resummation precisely cancels the logarithmic divergences of the perturbative corrections $\mathcal{A}$ and $\mathcal{B}$ in the limit $y\to0$, leaving only process-dependent constant and power-suppressed contributions. We note that at NNLO, the expansion of the NNLL resummation cancels all but a single-logarithmic term proportional to $\alphas^3L$ in $\mathcal{C}(y)$, which would require N$^3$LL resummation to be fully captured.
The NNLO coefficient $\mathcal{C}(y)$ then amends the cumulant by this term in addition to process-dependent contributions.
We have numerically cross checked the expansion of the NNLL result against the fixed-order calculation, finding excellent agreement up to NLO and the expected residual $\alphas^3 L$ dependence at NNLO.
In~\eqref{eq:matching}, the logarithms are modified as~\cite{Jones:2003yv}
\begin{equation}
  L \to L' = \log\left(\left(\frac{1}{y}\right) - \left(\frac{1}{y_\mathrm{max}}\right) + 1\right) \,,
  \label{eq:modifiedLog}
\end{equation}
to recover the physical behaviour of the cumulant distribution at large values of $y$, and ensure that $\Sigma(y_{\mathrm{max}})=1$.
We have determined the values of $y_\mathrm{max}$ analytically at LO and numerically from our fixed-order calculations at NLO and NNLO.

\begin{figure*}[t]
	\centering
	\hspace{-0.5cm}\includegraphics[width=0.3\textwidth]{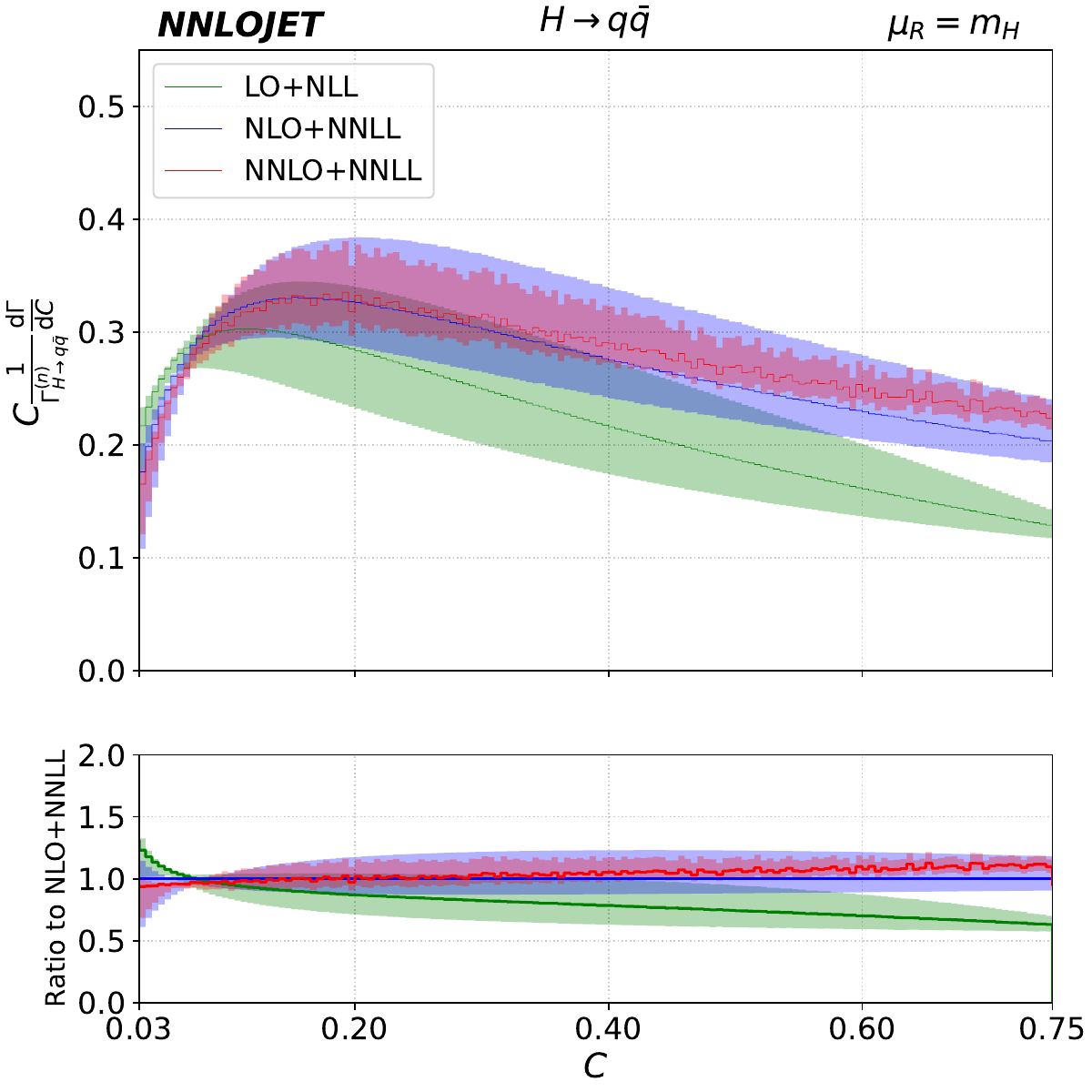}
	\hspace{0.3cm}
	\includegraphics[width=0.3\textwidth]{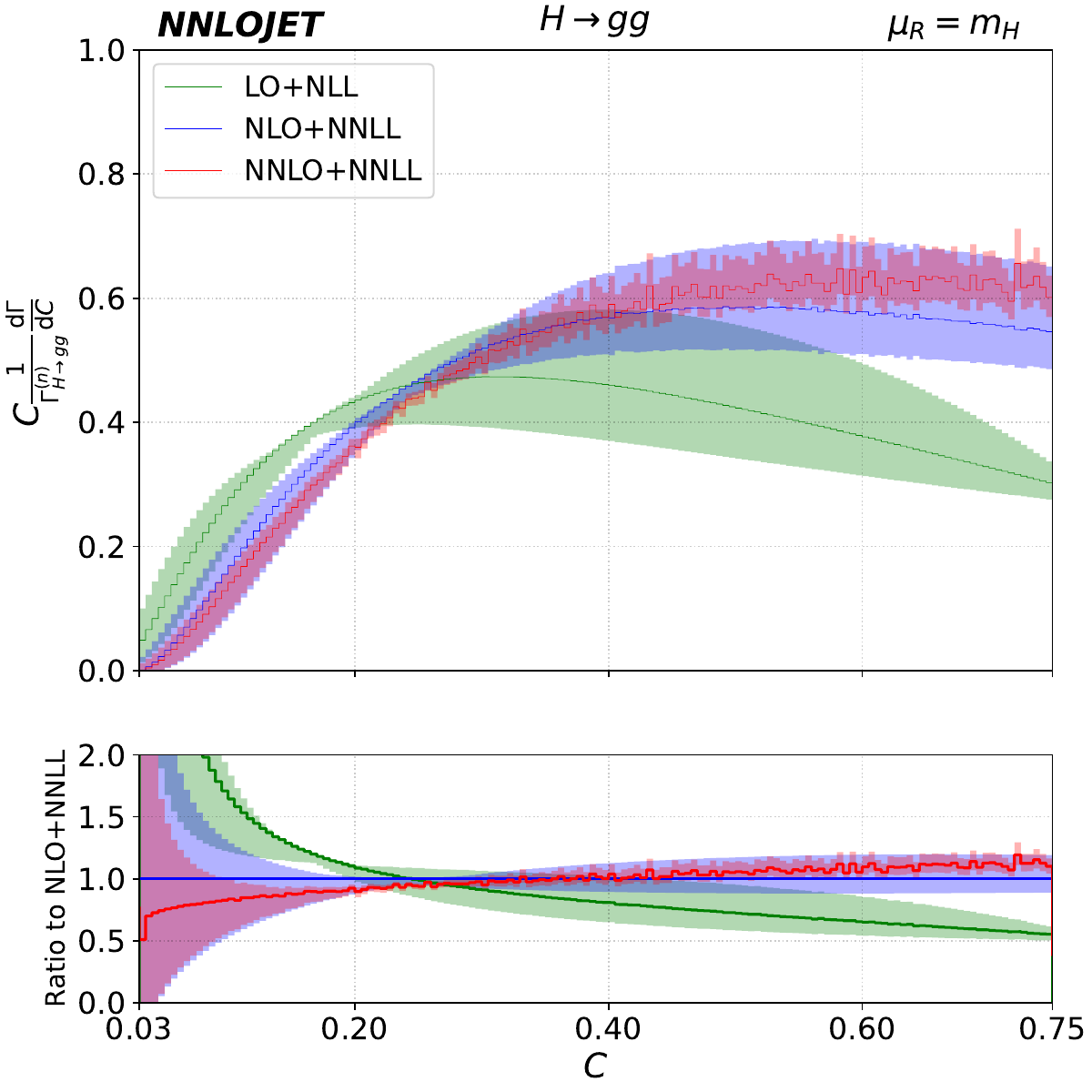}
	\hspace{0.3cm}
	\includegraphics[width=0.3\textwidth]{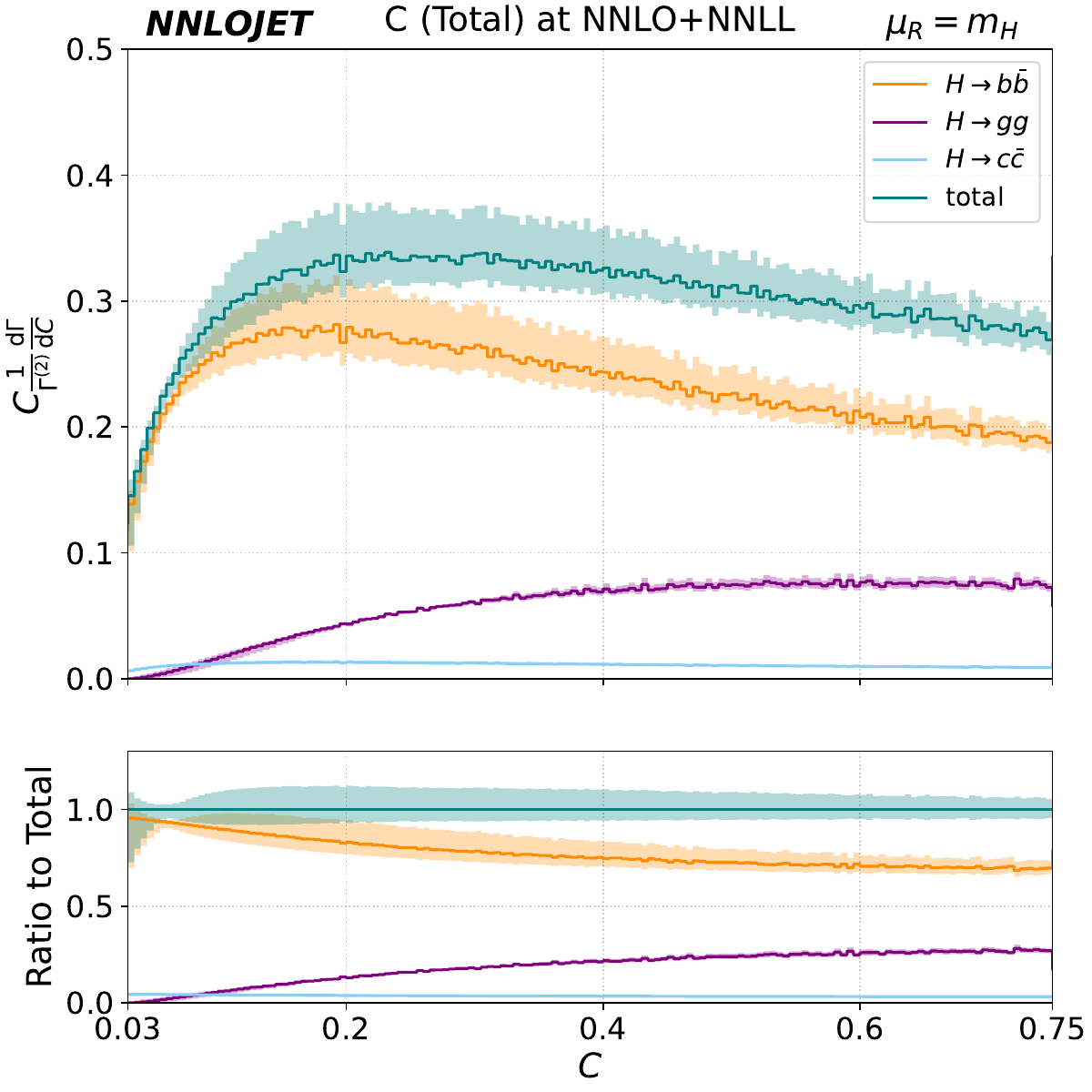}\\
	\hspace{-0.5cm}\includegraphics[width=0.3\textwidth]{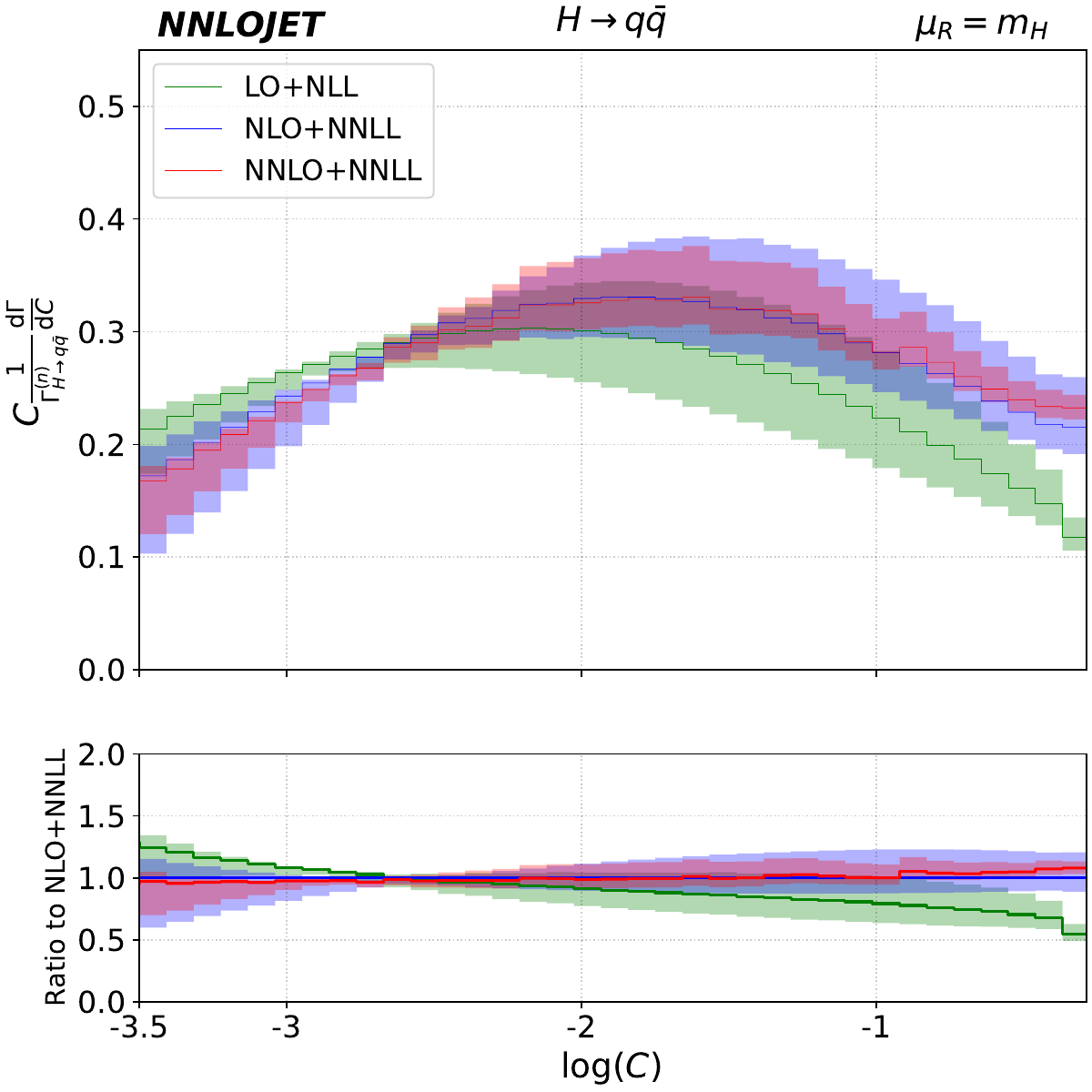}
	\hspace{0.3cm}
	\includegraphics[width=0.3\textwidth]{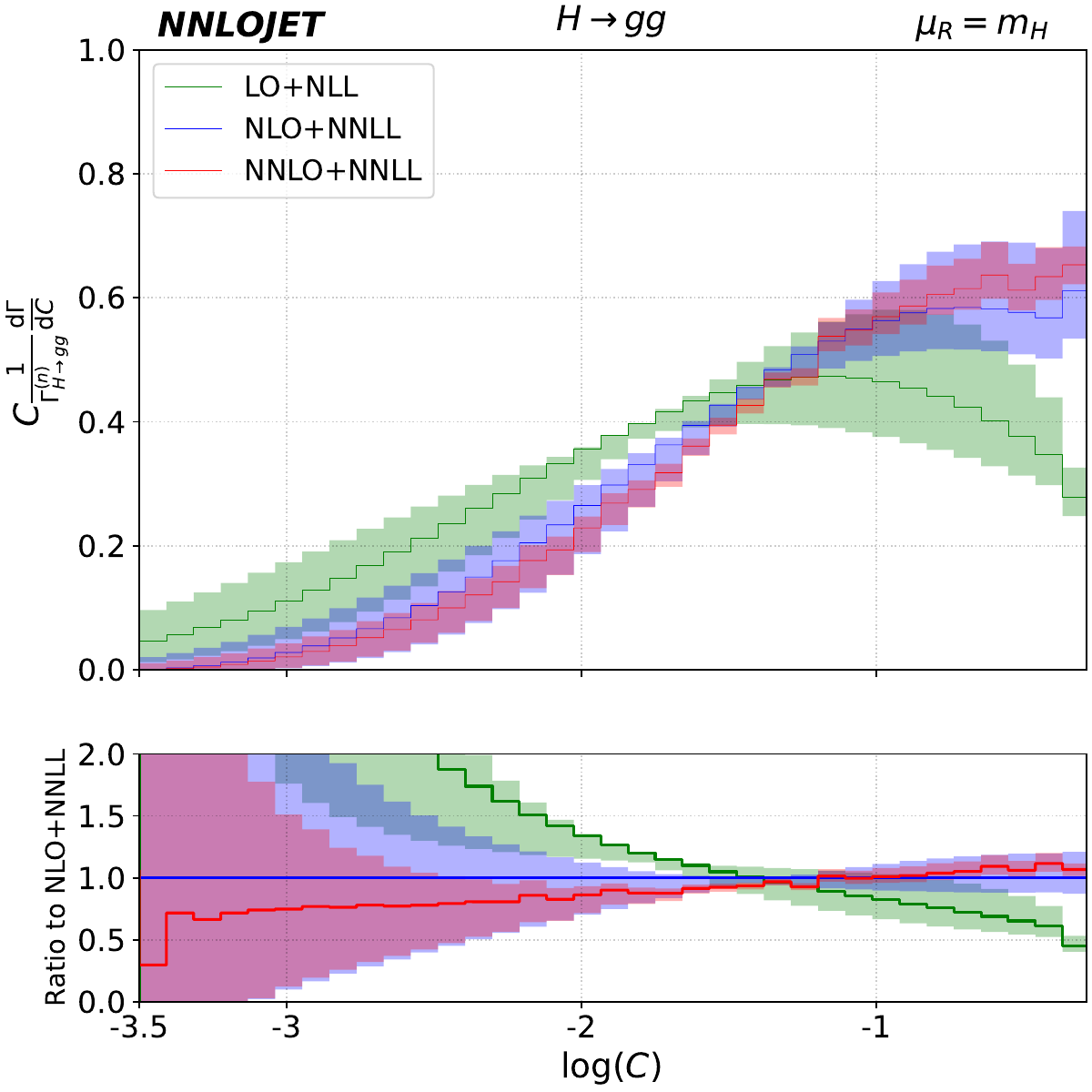}
	\hspace{0.3cm}
	\includegraphics[width=0.3\textwidth]{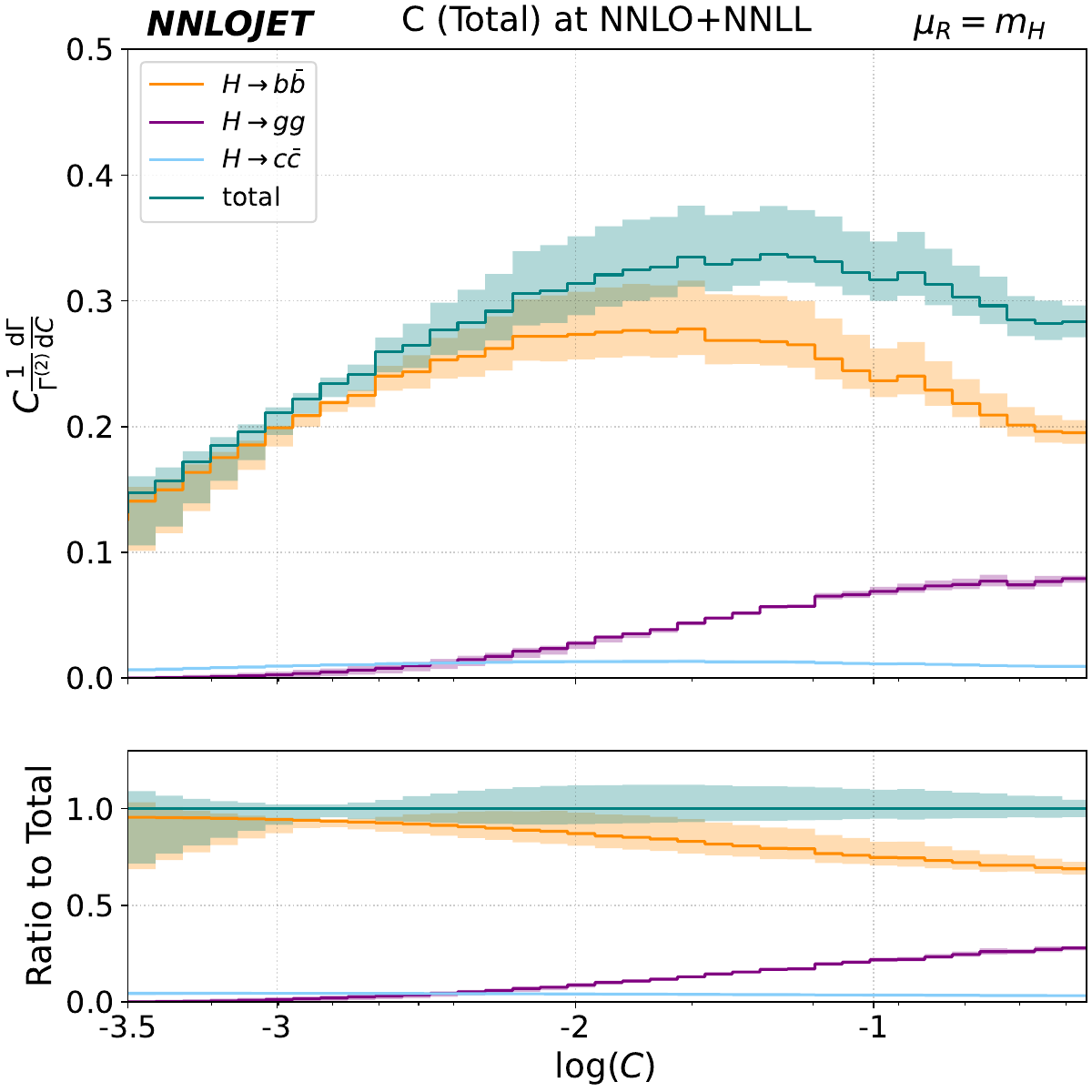}
	\caption{$C$-parameter distribution results for fixed order predictions matched to resummation in the logR scheme for the $H\to q\bar{q}$ channel (left column), $H\to gg$ channel (middle column) and the sum over all channels (right column) for both linear-binning (top row) and logarithmic-binning (bottom row). For the frames in the two leftmost columns, the curves represent the LO calculation matched to NLL resummation (green), NLO matched to NNLL (blue), and NNLO matched to NNLL (red), and the ratio to NLO+NNLL is shown in the lower frame. For the frames in the rightmost column, the curves represent the sum over all decay channels at NNLO matched to NNLL (teal), the $H\to b\bar{b}$ channel (orange), the $H\to gg$ channel (purple) and the $H \to c\bar{c}$ channel (light blue), and the ratio to the sum is shown in the lower frame.}
	\label{fig:C}
\end{figure*}

\section{Results}
\label{sec:results}
We present matched predictions for the heavy-jet mass and $C$-parameter distributions in the $H\to b\bar{b}$, $H\to c\bar{c}$, and $H\to gg$ decay modes at NNLO+NNLL.
We consider an on-shell Higgs boson with mass $m_{H}=125.09$ GeV, and work in the $G_{\mu}$ scheme with electroweak input parameters
\begin{equation}
  G_F = 1.1664\cdot 10^{-5}~\text{GeV}^{-2}, \quad m_Z = 91.200~\text{GeV}
\end{equation}
yielding a Higgs vacuum expectation value of $v=246.22~\GeV$. Our choice of renormalisation scale is $\mu_R=m_H$. To assess the uncertainties from missing higher-order contributions, we perform combined variations of both the renormalisation and resummation scales, $\mu_R\to x_{\mu}\mu_R$ and $y \to x_L y$ with \mbox{$x_{\mu},x_L\in [1/2,2]$}, around the central scales. 
The strong coupling is evolved from the input value \mbox{$\alpha_s(m_Z)=0.11800$}.
For the decay of a Higgs boson to quarks and gluons, we present differential distributions
\begin{eqnarray}
  \dfrac{\text{d}\Sigma_{H\to X}(y)}{\text{d}\log(y)}=y\dfrac{1}{\Gamma^{(k)}_{H\to X}(m_H,\mu_R)}\dfrac{\text{d}\Gamma_{H\to X}(m_H,\mu_R)}{\text{d} y},
\end{eqnarray}
both linearly- and logarithmically-binned, where $k$ indicates the perturbative order of the fixed-order prediction that the resummed result is matched to.
We also provide results for the sum of decay channels given by
\begin{equation}
  y\frac{1}{\Gamma^{(k)}(m_H,\mu_R)}\sum_{X}\frac{\rd \Gamma_{H\to X}(m_H,\mu_R)}{\rd y}, \quad\text{with}\,X=b\bar{b},c\bar{c},gg \,.
\end{equation}
with
\begin{equation}
  \Gamma^{(k)}=\Gamma^{(k)}_{H\to b\bar{b}}+\Gamma^{(k)}_{H\to c\bar{c}}+\Gamma^{(k)}_{H\to gg}\,.
\end{equation}
We rescale the effective coupling $\lambda_0(\mu^2)$ in~\eqref{eq:couplings} to include finite top, bottom, and charm mass effects \cite{Spira:1997dg}, as well as electroweak corrections \cite{Actis:2008ug}. 

We consider kinematically massless bottom and charm quarks, while keeping a non-vanishing Yukawa coupling. The bottom- and charm-quark Yukawa couplings are obtained using quark masses running in the $\overline{\text{MS}}$ scheme as $y_b(m_H)=m_b/v=0.011309$, \mbox{$y_c(m_H)=m_c/v=0.0024629$}, and we use the top-quark $\overline{\text{MS}}$ mass $m_t(m_H)=166.48$ GeV.
Quark masses entering the Yukawa couplings are evolved using the results of \cite{Vermaseren:1997fq}. 

In Fig.~\ref{fig:HJM} and~\ref{fig:C}, we show matched predictions for the heavy-jet mass and $C$-parameter, respectively, at LO+NLL, NLO+NNLL and NNLO+NNLL in $H\to q\bar{q}$ and $H\to gg$ decays, as well as results for the sum over all channels at NNLO+NNLL.
The first row shows linearly-binned results, while the second row shows plots on a logarithmic axis to enhance the resummation region.
We exclude the regions \mbox{$\rho_\mathrm{H}>1/3$} and \mbox{$C>3/4$} from the presented kinematical range, as the distributions exhibit a Sudakov shoulder there, which needs a dedicated resummed calculation to be properly described.

For the heavy-jet mass, shown in Fig.~\ref{fig:HJM}, we observe small differences between the NLO+NNLL and LO+NLL predictions in the resummation region $\rho_\mathrm{H} \lesssim 0.15$ in both decay channels.
Larger negative corrections up to -25\% are found at NLO+NNLL for $\rho_\mathrm{H} > 0.15$ in both decay channels.
Similarly, the inclusion of the NNLO correction leads to relevant effects only for $\rho_\mathrm{H} \gtrsim 0.25$ for the $H\to q\bar{q}$ mode and for $\rho_\mathrm{H}\gtrsim0.2$ for the $H\to gg$ mode.
In general, the inclusion of higher-logarithmic and higher-order corrections yields visibly smaller scale-variation bands, and the theory uncertainties of lower orders contain higher-order corrections in the bulk of the distributions, indicating good convergence of the matched predictions, especially for the Yukawa mode.
This agrees with the observations at fixed order in~\cite{Fox:2025qmp} for intermediate to high values of $\rho_\mathrm{H}$.

For the $C$-parameter distribution, we see larger differences between NLO+NNLL and LO+NLL predictions, with positive corrections for values of $C\gtrsim0.05$ in $H\to q\bar{q}$ and $C\gtrsim0.25$ in $H\to gg$, and negative corrections in the complementary regions.
For $C\gtrsim0.55$ in both decay modes, the LO+NLL and NLO+NNLL uncertainty bands do not overlap.
The inclusion of NNLO corrections leads to an enhancement for intermediate and high values of $C$, reaching up to $+15\%$ before the Sudakov shoulder at $C=0.75$, which is fully captured by the NLO+NNLL scale variation bands.
In the low $C$ region, the effect of the NNLO correction is small for the Yukawa mode, similarly to what is observed for the heavy-jet mass and thrust distributions~\cite{Fox:2025qmp}.
On the other hand, it leads to a sizeable modification of the distribution in the gluonic mode.
While both NLO and NNLO predictions are matched to NNLL-accurate resummation, in the NNLO+NNLL calculation, as mentioned above, there is an unmatched single logarithm proportional to $\alpha_s^3L$, which is particularly visible for low values of the $C$-parameter in the $H\to gg$ channel, as can be inferred from the middle plot in the lower row of Fig.~\ref{fig:C}.
An N$^3$LL-accurate matched calculation is needed to account for this term.
Moreover, the peak of the $H\to gg$ distribution is shifted close to the Sudakov shoulder in the NNLO+NNLL predictions, indicating that a common resummation of dijet logarithms and Sudakov-shoulder logarithms is required in this case.
We have checked that this is mainly due to higher-order corrections in the fixed-order contribution and is in line with a similar shift observed when going from LO+NLL to NLO+NLL in \cite{Gehrmann-DeRidder:2024avt}.

Considering the rightmost panels of Fig.~\ref{fig:HJM} and ~\ref{fig:C}, for both event shapes we observe that the relative fractions of the different decay modes for the matched predictions largely resemble the ones obtained in the pure fixed-order calculation~\cite{Fox:2025qmp}.
In particular, the relative contributions are roughly constant in the considered range, with the gluonic mode decreasing below $\rho_\mathrm{H}\approx 0.05$ and $C\approx 0.35$.
As expected, the inclusion of resummation effects prevents the predictions from becoming negative, hence unphysical, as $\rho_\mathrm{H},C\to 0$.

\section{Conclusions}
\label{sec:conclusions}
We presented the resummation of logarithmically-enhanced corrections in the two-particle limit for the heavy-jet mass and $C$-parameter event shapes in hadronic decays of the Higgs boson to quarks and gluons.
The calculation is performed analytically up to NNLL relying on the \ARES~\cite{Banfi:2014sua,Banfi:2018mcq} formalism, and closely follows the procedure for thrust, described in~\cite{Fox:2025txz}.
Resummed predictions have been matched to the NNLO-accurate fixed-order results that we obtained in~\cite{Fox:2025qmp} in the logR scheme.
We illustrated the behaviour of matched calculations separately for the Yukawa-induced $H\to q\bar{q}$ channel and for the gluonic decay mode $H\to gg$, mediated by an effective $Hgg$ interaction.
Moreover, we presented phenomenologically viable results for the sum over all relevant hadronic decays of the Higgs boson and highlighted the relative contribution of the $H\to b\bar{b}$, $H\to c\bar{c}$ and $H\to gg$ channels differentially in the considered observables.
The inclusion of NNLL corrections leads to significantly reduced uncertainty bands in the bulk of the distributions.
Both, higher-logarithmic and higher-order corrections are small for the heavy-jet mass in both Higgs-decay modes.
While the same is true for NNLO corrections in the $C$-parameter, NNLL corrections are large in this case, with a significant shift of the peak of the distribution towards the Sudakov shoulder in $H\to gg$ decays.

This work fits within recent efforts~\cite{Alioli:2020fzf,Coloretti:2022jcl,Gao:2016jcm,Gao:2019mlt,Gao:2020vyx,Gehrmann-DeRidder:2023uld,Gehrmann-DeRidder:2024avt,Fox:2025cuz,Fox:2025qmp,Fox:2025txz,Ju:2023dfa,Knobbe:2023njd,Luo:2019nig,Mo:2017gzp,vanBeekveld:2024wws,Yan:2023xsd,Zhu:2023oka} aimed at the precise description of hadronic final states in Higgs boson decays, where a fully consistent description \cite{Caletti:2025pot} will also have to account for the bosst of the Higgs boson in the $e^+e^-$ collider frame and for the accompanying $Z$ boson decay products.
Moreover, our results offer important insights about the different features of resummed calculations in the presence of quark or gluon hard radiators.
In particular, we observe, as expected, an enhanced sensitivity to logarithmic corrections in the gluonic mode, manifested as larger theory uncertainties and a more visible impact of unmatched N$^3$LL logarithms, especially for the $C$-parameter.
To obtain a reliable prediction of the $C$-parameter distribution in $H\to gg$ decays, our NNLO+NNLL results should be amended by Sudakov-shoulder resummation in the future.
The resummation of Sudakov-shoulder logarithms in the heavy-jet mass distribution will extend the region where our results are reliable towards the multi-particle region, but is less pressing than in the case of the $C$-parameter.
We expect to extend our framework to include NNLO+NNLL accurate predictions for the full set of classical event-shape observables in Higgs decays to quarks and gluons in the future.
For both event-shape observables considered in this letter, we anticipate that an extension to N$^3$LL resummation is within reach in the near future.

\section*{Acknowledgements}
AG acknowledges the support of the Swiss National Science Foundation (SNF) under contract 200021-231259 and of the Swiss National Supercomputing Centre (CSCS) under project ID ETH5f. 
TG has received funding from the Swiss National Science Foundation (SNF) under contract 240015 and from the European Research Council (ERC) under the European Union's Horizon 2020 research and innovation programme grant agreement 101019620 (ERC Advanced Grant TOPUP).
NG gratefully acknowledges support from the UK Science and Technology Facilities Council (STFC) under contract ST/X000745/1 and hospitality from the Pauli Center for Theoretical Studies, Zurich. 
Part of the computations were carried out on the PLEIADES cluster at the University of Wuppertal, supported by the Deutsche Forschungsgemeinschaft (DFG, grant No. INST 218/78-1 FUGG) and the Bundesministerium f{\"u}r Bildung und Forschung (BMBF).
MM is supported by a Royal Society Newton International Fellowship (NIF/R1/232539).

\appendix
\section{Fixed-order expansion coefficients}\label{app:Gij}
Here, we collect the fixed-order expansion coefficients up to $\mathcal{O}(\alphas^3)$ for both Higgs-decay categories, $H\to q\bar{q}$ and $H\to gg$.

\subsection{Heavy-jet mass}
The fixed-order expansion of the resummation yields the following coefficients for quark-antiquark radiators,
\allowdisplaybreaks
\begin{align}
  G_{12,q\bar{q}}&= -2\CF\,,\\
  G_{23,q\bar{q}}&= -\frac{11}{3}\CA\CF + \frac{2}{3}\NF\,,\\
  G_{34,q\bar{q}}&= -\frac{847}{108}\CA^2\CF + \frac{77}{27}\CA\CF\NF - \frac{7}{27}\CF\NF^2\,,\\
  G_{11,q\bar{q}}&= 3\CF\,,\\
  G_{22,q\bar{q}}&= \left(-\frac{169}{36} + \frac{\pi^2}{3}\right)\CA\CF - \frac{2\pi^2}{3}\CF^2 + \frac{11}{18}\CF\NF\,,\\
  \begin{split}
    G_{33,q\bar{q}}&= \left(-\frac{3197}{108} + \frac{11\pi^2}{9}\right)\CA^2\CF - \frac{11\pi^2}{3}\CA\CF^2\\
    &\quad + \frac{16\zeta_3}{3}\CF^3 + \left(\frac{256}{27} - \frac{2\pi^2}{9}\right)\CA\CF\NF\\
    &\quad + \left(1 + \frac{2\pi^2}{3}\right)\CF^2\NF - \frac{17}{27}\CF\NF^2\,,
  \end{split}\\
  G_{21,q\bar{q}}&= \left(\frac{57}{4} - 6\zeta_3\right)\CA\CF + \left(\frac{3}{4} + 4\zeta_3\right)\CF^2 - \frac{5}{2}\CF\NF\,,\\
  \begin{split}
    G_{32,q\bar{q}}&= \left(-\frac{11323}{648} + \frac{85\pi^2}{18} + 11\zeta_3 - \frac{11\pi^4}{90}\right)\CA^2\CF \\
    &\quad + \left(\frac{11}{8} - \frac{239\pi^2}{108} - 44\zeta_3 + \frac{2\pi^4}{9}\right)\CA\CF^2 \\
    &\quad + \left(-12\zeta_3 + \frac{2\pi^4}{45}\right)\CF^3\\
    &\quad + \left(\frac{673}{324} - \frac{32\pi^2}{27} + 2\zeta_3\right)\CA\CF\NF \\
    &\quad + \left(\frac{43}{12} + \frac{13\pi^2}{54} + 4\zeta_3\right)\CF^2\NF \\
    &\quad + \left(\frac{35}{162} + \frac{2\pi^2}{27}\right)\CF\NF^2\,,
  \end{split}
\end{align}
and the following expansion coefficients for gluon-gluon radiators,
\begin{align}
  G_{12,gg}&= -2\CA\,,\\
  G_{23,gg}&= -\frac{11}{3}\CA^2 + \frac{2}{3}\CA\NF\,,\\
  G_{34,gg}&= -\frac{847}{108}\CA^3 + \frac{77}{27}\CA^2\NF - \frac{7}{27}\CA\NF^2\,,\\
  G_{11,gg}&= \frac{11}{3}\CA - \frac{2}{3}\NF\,,\\
  G_{22,gg}&= -\left(\frac{49}{12} - \frac{\pi^2}{3}\right)\CA^2 - \frac{1}{9}\CA\NF + \frac{1}{9}\NF^2\,,\\
  \begin{split}
    G_{33,gg}&= \left(-\frac{9349}{324} - \frac{22\pi^2}{9} + \frac{16\zeta_3}{3}\right)\CA^3 \\
    &\quad + \left(\frac{457}{54} + \frac{4\pi^2}{9}\right)\CA^2\NF \\
    &\quad + \CA\CF\NF - \frac{1}{3}\CA\NF^2 - \frac{2}{81}\NF^3\,,
  \end{split}\\
  \begin{split}
    G_{21,gg}&= \left(\frac{1025}{54} - 2\zeta_3\right)\CA^2 - \frac{158}{27}\CA\NF\\
    &\quad - \CF\NF + \frac{10}{27}\NF^2\,,
  \end{split}\\
  \begin{split}
    G_{32,gg}&= \left(-\frac{2545}{324} + \frac{337\pi^2}{108} - \frac{143\zeta_3}{3} + \frac{13\pi^4}{90}\right)\CA^3 \\
    &\quad + \left(-\frac{2225}{324} - \frac{5\pi^2}{3} + \frac{38\zeta_3}{3}\right)\CA^2\NF \\
    &\quad + \left(\frac{11}{6} - 4\zeta_3\right)\CA\CF\NF\\
    &\quad + \left(\frac{371}{162} + \frac{5\pi^2}{27}\right)\CA\NF^2\\
    &\quad + \frac{1}{2}\CF\NF^2 - \frac{10}{81}\NF^3\,,
  \end{split}
\end{align}
The first-order hard coefficients are identical to the ones in thrust and are given by
\begin{align}
  c_{1}^{(H\to q\bar{q})} &= \left(-\frac{5}{2} + \frac{\pi^2}{3}\right)\CF\,,\\
  c_{1}^{(H\to gg)} &= \left(-\frac{85}{18} + \frac{\pi^2}{3}\right)\CA + \frac{11}{9}\NF\,.
\end{align}

\subsection{$C$-parameter}
The fixed-order expansion expansion coefficients of the $C$-parameter are given by
\begin{align}
  G_{12} &= \bar{G}_{12} \,,\\
  G_{23} &= \bar{G}_{23} \,,\\
  G_{34} &= \bar{G}_{34} \,,\\
  G_{11} &= \bar{G}_{11} + 2\bar{G}_{12}\log(6) \,,\\
  G_{22} &= \bar{G}_{22} + 3\bar{G}_{23}\log(6) \,,\\
  G_{33} &= \bar{G}_{33} + 4\bar{G}_{34}\log(6) \,,\\
  G_{21} &= \bar{G}_{21} + 2\bar{G}_{22}\log(6) + 3\bar{G}_{23}\log(6)^2 \,,\\
  G_{32} &= \bar{G}_{32} + 3\bar{G}_{33}\log(6) + 6\bar{G}_{34}\log(6)^2 \,,
\end{align}
in terms of the following expansion coefficients of the observable $y=C/6$~\cite{Catani:1998sf} for quark-antiquark radiators,
\begin{align}
  \bar{G}_{12,q\bar{q}}&= -2\CF\,,\\
  \bar{G}_{23,q\bar{q}}&= -\frac{11}{3}\CA\CF + \frac{2}{3}\CF\NF\,,\\
  \bar{G}_{34,q\bar{q}}&= -\frac{847}{108}\CA^2\CF + \frac{77}{27}\CA\CF\NF - \frac{7}{27}\CF\NF^2\,,\\
  \bar{G}_{11,q\bar{q}}&= 3\CF\,,\\
  \bar{G}_{22,q\bar{q}}&= \left(-\frac{169}{36} + \frac{\pi^2}{3}\right)\CA\CF - \frac{4\pi^2}{3}\CF^2 + \frac{11}{18}\CF\NF\,,\\
  \begin{split}
    \bar{G}_{33,q\bar{q}}&= \left(-\frac{3197}{108} + \frac{11\pi^2}{9}\right)\CA^2\CF - \frac{22\pi^2}{3}\CA\CF^2 \\
    &\quad + \frac{64\zeta_3}{3}\CF^3 + \left(\frac{256}{27} - \frac{2\pi^2}{9}\right)\CA\CF\NF\\
    &\quad + \left(1 + \frac{4\pi^2}{3}\right)\CF^2\NF - \frac{17}{27}\CF\NF^2\,,
  \end{split}\\
  \begin{split}
    \bar{G}_{21,q\bar{q}}&= \left(\frac{57}{4} + \frac{11\pi^2}{9} - 6\zeta_3\right)\CA\CF \\
    &\quad + \left(\frac{3}{4} + \pi^2 - 4\zeta_3\right)\CF^2 + \left(-\frac{5}{2} - \frac{2\pi^2}{9}\right)\CF\NF\,,
  \end{split}\\
  \begin{split}
    \bar{G}_{32,q\bar{q}}&= \left(-\frac{11323}{648} + \frac{497\pi^2}{54} + 11\zeta_3 - \frac{11\pi^4}{90}\right)\CA^2\CF\\
    &\quad + \left(\frac{11}{8} - \frac{70\pi^2}{27} - 110\zeta_3 + \frac{4\pi^4}{9}\right)\CA\CF^2\\
    &\quad + \left(-48\zeta_3 + \frac{8\pi^4}{45}\right)\CF^3 \\
    &\quad + \left(\frac{673}{324} - \frac{76\pi^2}{27} + 2\zeta_3\right)\CA\CF\NF \\
    &\quad + \left(\frac{43}{12} + \frac{4\pi^2}{27} + 16\zeta_3\right)\CF^2\NF \\
    &\quad + \left(\frac{35}{162} + \frac{2\pi^2}{9}\right)\CF\NF^2\,,
  \end{split}
\end{align}
and for gluon-gluon radiators,
\begin{align}
  \bar{G}_{12,gg}&= -2\CA\,,\\
  \bar{G}_{23,gg}&= -\frac{11}{3}\CA^2 + \frac{2}{3}\CA\NF\,,\\
  \bar{G}_{34,gg}&= -\frac{847}{108}\CA^3 + \frac{77}{27}\CA^2\NF - \frac{7}{27}\CA\NF^2 \,,\\
  \bar{G}_{11,gg}&= \frac{11}{3}\CA - \frac{2}{3}\NF\,,\\
  \bar{G}_{22,gg}&= \left(-\frac{49}{12} - \pi^2\right)\CA^2 - \frac{1}{9}\CA\NF + \frac{1}{9}\NF^2\,,\\
  \begin{split}
    \bar{G}_{33,gg}&= \left(-\frac{9349}{324} - \frac{55\pi^2}{9} + \frac{64\zeta_3}{3}\right)\CA^3\\
    &\quad + \left(\frac{457}{54} + \frac{10\pi^2}{9}\right)\CA^2\NF\\
    &\quad + \CA\CF\NF - \frac{1}{3}\CA\NF^2 - \frac{2}{81}\NF^3\,,
  \end{split}\\
  \begin{split}
    \bar{G}_{21,gg}&= \left(\frac{1025}{54} + \frac{22\pi^2}{9} - 10\zeta_3\right)\CA^2\\
    &\quad + \left(-\frac{158}{27} - \frac{4\pi^2}{9}\right)\CA\NF - \CF\NF + \frac{10}{27}\NF^2\,,
  \end{split}\\
  \begin{split}
    \bar{G}_{32,gg}&= \left(-\frac{2545}{324} + \frac{445\pi^2}{54} - \frac{473\zeta_3}{3} + \frac{\pi^4}{2}\right)\CA^3\\
    &\quad + \left(-\frac{2225}{324} - \frac{124\pi^2}{27} + \frac{98\zeta_3}{3}\right)\CA^2\NF\\
    &\quad + \left(\frac{11}{6} - 4\zeta_3\right)\CA\CF\NF \\
    &\quad + \left(\frac{371}{162} + \frac{14\pi^2}{27}\right)\CA\NF^2\\
    &\quad + \frac{1}{2}\CF\NF^2 - \frac{10}{81}\NF^3\,.
  \end{split}
\end{align}
The first-order hard coefficients $c_1^{(H\to X)}$ for the $C$-parameter are given by
\begin{equation}
  c_1^{(H\to X)} = \bar{c}_1^{(H\to X)}\bar{G}_{11}\log(6) + \bar{G}_{12}\log(6)^2
\end{equation}
in terms of the respective expansion coefficients $\bar{G}_{12}$ and $\bar{G}_{11}$ above and the hard coefficients $\bar{c}_1^{(H\to X)}$ for the observable \mbox{$y=C/6$},
\begin{align}
  \bar{c}_{1}^{(H\to q\bar{q})} &= \left(-\frac{5}{2} + \frac{2\pi^2}{3}\right)\CF\,,\\
  \bar{c}_{1}^{(H\to gg)} &= \left(-\frac{85}{18} + \frac{2\pi^2}{3}\right)\CA + \frac{11}{9}\NF\,.
\end{align}

\bibliographystyle{elsarticle-harv} 
\bibliography{main}

@article{Gehrmann-DeRidder:2005btv,
    author = "Gehrmann-De Ridder, A. and Gehrmann, T. and Glover, E. W. Nigel",
    title = "{Antenna subtraction at NNLO}",
    eprint = "hep-ph/0505111",
    archivePrefix = "arXiv",
    reportNumber = "ZU-TH-07-05, IPPP-05-18",
    doi = "10.1088/1126-6708/2005/09/056",
    journal = "JHEP",
    volume = "09",
    pages = "056",
    year = "2005"
}

@article{Gehrmann-DeRidder:2007vsv,
    author = "Gehrmann-De Ridder, A. and Gehrmann, T. and Glover, E. W. N. and Heinrich, G.",
    title = "{NNLO corrections to event shapes in $e^+e^-$ annihilation}",
    eprint = "0711.4711",
    archivePrefix = "arXiv",
    primaryClass = "hep-ph",
    reportNumber = "ZU-TH-27-07, IPPP-07-90, EDINBURGH-2007-47",
    doi = "10.1088/1126-6708/2007/12/094",
    journal = "JHEP",
    volume = "12",
    pages = "094",
    year = "2007"
}

@article{Mondini:2019gid,
    author = "Mondini, Roberto and Schiavi, Matthew and Williams, Ciaran",
    title = "{N$^{3}$LO predictions for the decay of the Higgs boson to bottom quarks}",
    eprint = "1904.08960",
    archivePrefix = "arXiv",
    primaryClass = "hep-ph",
    doi = "10.1007/JHEP06(2019)079",
    journal = "JHEP",
    volume = "06",
    pages = "079",
    year = "2019"
}

@article{Mondini:2019vub,
    author = "Mondini, Roberto and Williams, Ciaran",
    title = "{$ H\to b\overline{b}j $ at next-to-next-to-leading order accuracy}",
    eprint = "1904.08961",
    archivePrefix = "arXiv",
    primaryClass = "hep-ph",
    doi = "10.1007/JHEP06(2019)120",
    journal = "JHEP",
    volume = "06",
    pages = "120",
    year = "2019"
}

@article{Gao:2019mlt,
    author = "Gao, Jun and Gong, Yinqiang and Ju, Wan-Li and Yang, Li Lin",
    title = "{Thrust distribution in Higgs decays at the next-to-leading order and beyond}",
    eprint = "1901.02253",
    archivePrefix = "arXiv",
    primaryClass = "hep-ph",
    doi = "10.1007/JHEP03(2019)030",
    journal = "JHEP",
    volume = "03",
    pages = "030",
    year = "2019"
}

@article{Anastasiou:2011qx,
    author = "Anastasiou, Charalampos and Herzog, Franz and Lazopoulos, Achilleas",
    title = "{The fully differential decay rate of a Higgs boson to bottom-quarks at NNLO in QCD}",
    eprint = "1110.2368",
    archivePrefix = "arXiv",
    primaryClass = "hep-ph",
    doi = "10.1007/JHEP03(2012)035",
    journal = "JHEP",
    volume = "03",
    pages = "035",
    year = "2012"
}

@article{DelDuca:2015zqa,
    author = "Del Duca, Vittorio and Duhr, Claude and Somogyi, G\'abor and Tramontano, Francesco and Tr\'ocs\'anyi, Zolt\'an",
    title = "{Higgs boson decay into b-quarks at NNLO accuracy}",
    eprint = "1501.07226",
    archivePrefix = "arXiv",
    primaryClass = "hep-ph",
    reportNumber = "CERN-PH-TH-2015-005, CP3-14-82",
    doi = "10.1007/JHEP04(2015)036",
    journal = "JHEP",
    volume = "04",
    pages = "036",
    year = "2015"
}

@article{Clavelli:1981yh,
    author = "Clavelli, L. and Wyler, D.",
    title = "{Kinematical Bounds on Jet Variables and the Heavy Jet Mass Distribution}",
    reportNumber = "BONN-HE-81-3",
    doi = "10.1016/0370-2693(81)90248-3",
    journal = "Phys. Lett. B",
    volume = "103",
    pages = "383--387",
    year = "1981"
}

@article{Parisi:1978eg,
    author = "Parisi, G.",
    title = "{Super Inclusive Cross-Sections}",
    reportNumber = "LPTENS 78/5",
    doi = "10.1016/0370-2693(78)90061-8",
    journal = "Phys. Lett. B",
    volume = "74",
    pages = "65--67",
    year = "1978"
}

@article{Donoghue:1979vi,
    author = "Donoghue, John F. and Low, F. E. and Pi, So-Young",
    title = "{Tensor Analysis of Hadronic Jets in Quantum Chromodynamics}",
    reportNumber = "MIT-CTP-771",
    doi = "10.1103/PhysRevD.20.2759",
    journal = "Phys. Rev. D",
    volume = "20",
    pages = "2759",
    year = "1979"
}

@article{Alioli:2020fzf,
    author = "Alioli, Simone and Broggio, Alessandro and Gavardi, Alessandro and Kallweit, Stefan and Lim, Matthew A. and Nagar, Riccardo and Napoletano, Davide and Rottoli, Luca",
    title = "{Resummed predictions for hadronic Higgs boson decays}",
    eprint = "2009.13533",
    archivePrefix = "arXiv",
    primaryClass = "hep-ph",
    doi = "10.1007/JHEP04(2021)254",
    journal = "JHEP",
    volume = "04",
    pages = "254",
    year = "2021"
}

@article{Gao:2016jcm,
    author = "Gao, Jun",
    title = "{Probing light-quark Yukawa couplings via hadronic event shapes at lepton colliders}",
    eprint = "1608.01746",
    archivePrefix = "arXiv",
    primaryClass = "hep-ph",
    doi = "10.1007/JHEP01(2018)038",
    journal = "JHEP",
    volume = "01",
    pages = "038",
    year = "2018"
}

@article{Spira:1997dg,
	author = "Spira, Michael",
	title = "{QCD effects in Higgs physics}",
	eprint = "hep-ph/9705337",
	archivePrefix = "arXiv",
	reportNumber = "CERN-TH-97-068, CERN-TH-97-68",
	doi = "10.1002/(SICI)1521-3978(199804)46:3<203::AID-PROP203>3.0.CO;2-4",
	journal = "Fortsch. Phys.",
	volume = "46",
	pages = "203--284",
	year = "1998"
}

@article{Actis:2008ug,
    author = "Actis, Stefano and Passarino, Giampiero and Sturm, Christian and Uccirati, Sandro",
    title = "{NLO Electroweak Corrections to Higgs Boson Production at Hadron Colliders}",
    eprint = "0809.1301",
    archivePrefix = "arXiv",
    primaryClass = "hep-ph",
    reportNumber = "PITHA-08-20, SFB-CPP-08-62, TTP08-38",
    doi = "10.1016/j.physletb.2008.10.018",
    journal = "Phys. Lett. B",
    volume = "670",
    pages = "12--17",
    year = "2008"
}

@article{ALEPH:2003obs,
    author = "Heister, A. and others",
    collaboration = "ALEPH",
    title = "{Studies of QCD at $e^+e^-$ centre-of-mass energies between $91~\mathrm{GeV}$ and $209~\mathrm{GeV}$}",
    reportNumber = "CERN-EP-2003-084",
    doi = "10.1140/epjc/s2004-01891-4",
    journal = "Eur. Phys. J. C",
    volume = "35",
    pages = "457--486",
    year = "2004"
}

@article{DELPHI:2003yqh,
    author = "Abdallah, J. and others",
    collaboration = "DELPHI",
    title = "{A Study of the energy evolution of event shape distributions and their means with the DELPHI detector at LEP}",
    eprint = "hep-ex/0307048",
    archivePrefix = "arXiv",
    reportNumber = "CERN-EP-2002-082",
    doi = "10.1140/epjc/s2003-01198-0",
    journal = "Eur. Phys. J. C",
    volume = "29",
    pages = "285--312",
    year = "2003"
}

@article{Banfi:2014sua,
    author = "Banfi, Andrea and McAslan, Heather and Monni, Pier Francesco and Zanderighi, Giulia",
    title = "{A general method for the resummation of event-shape distributions in $e^+e^-$ annihilation}",
    eprint = "1412.2126",
    archivePrefix = "arXiv",
    primaryClass = "hep-ph",
    reportNumber = "OUTP-14-18P",
    doi = "10.1007/JHEP05(2015)102",
    journal = "JHEP",
    volume = "05",
    pages = "102",
    year = "2015"
}

@article{Mo:2017gzp,
    author = "Mo, Jonathan and Tackmann, Frank J. and Waalewijn, Wouter J.",
    title = "{A case study of quark-gluon discrimination at NNLL' in comparison to parton showers}",
    eprint = "1708.00867",
    archivePrefix = "arXiv",
    primaryClass = "hep-ph",
    reportNumber = "DESY-17-111, NIKHEF-2017-031",
    doi = "10.1140/epjc/s10052-017-5365-9",
    journal = "Eur. Phys. J. C",
    volume = "77",
    pages = "770",
    year = "2017"
}

@article{Catani:1992ua,
    author = "Catani, S. and Trentadue, L. and Turnock, G. and Webber, B. R.",
    title = "{Resummation of large logarithms in $e^+e^-$ event shape distributions}",
    reportNumber = "CERN-TH-6640-92, CAVENDISH-HEP-91-11",
    doi = "10.1016/0550-3213(93)90271-P",
    journal = "Nucl. Phys. B",
    volume = "407",
    pages = "3--42",
    year = "1993"
}

@article{Herzog:2017dtz,
    author = "Herzog, F. and Ruijl, B. and Ueda, T. and Vermaseren, J. A. M. and Vogt, A.",
    title = "{On Higgs decays to hadrons and the R-ratio at N$^{4}$LO}",
    eprint = "1707.01044",
    archivePrefix = "arXiv",
    primaryClass = "hep-ph",
    reportNumber = "NIKHEF-2017-029, LTH-1136",
    doi = "10.1007/JHEP08(2017)113",
    journal = "JHEP",
    volume = "08",
    pages = "113",
    year = "2017"
}

@article{Luo:2019nig,
    author = "Luo, Ming-Xing and Shtabovenko, Vladyslav and Yang, Tong-Zhi and Zhu, Hua Xing",
    title = "{Analytic Next-To-Leading Order Calculation of Energy-Energy Correlation in Gluon-Initiated Higgs Decays}",
    eprint = "1903.07277",
    archivePrefix = "arXiv",
    primaryClass = "hep-ph",
    doi = "10.1007/JHEP06(2019)037",
    journal = "JHEP",
    volume = "06",
    pages = "037",
    year = "2019"
}

@article{Gao:2020vyx,
    author = "Gao, Jun and Shtabovenko, Vladyslav and Yang, Tong-Zhi",
    title = "{Energy-energy correlation in hadronic Higgs decays: analytic results and phenomenology at NLO}",
    eprint = "2012.14188",
    archivePrefix = "arXiv",
    primaryClass = "hep-ph",
    reportNumber = "P3H-20-060, TTP20-035, ZU-TH 18/20",
    doi = "10.1007/JHEP02(2021)210",
    journal = "JHEP",
    volume = "02",
    pages = "210",
    year = "2021"
}

@article{Catani:1998sf,
    author = "Catani, S. and Webber, B. R.",
    title = "{Resummed $C$ parameter distribution in $e^+e^-$ annihilation}",
    eprint = "hep-ph/9801350",
    archivePrefix = "arXiv",
    reportNumber = "CERN-TH-98-14, CAVENDISH-HEP-97-16",
    doi = "10.1016/S0370-2693(98)00359-1",
    journal = "Phys. Lett. B",
    volume = "427",
    pages = "377--384",
    year = "1998"
}

@article{Vermaseren:1997fq,
    author = "Vermaseren, J. A. M. and Larin, S. A. and van Ritbergen, T.",
    title = "{The four loop quark mass anomalous dimension and the invariant quark mass}",
    eprint = "hep-ph/9703284",
    archivePrefix = "arXiv",
    reportNumber = "UM-TH-97-03, NIKHEF-97-012",
    doi = "10.1016/S0370-2693(97)00660-6",
    journal = "Phys. Lett. B",
    volume = "405",
    pages = "327--333",
    year = "1997"
}

@article{Fox:2025cuz,
    author = "Fox, Elliot and Gehrmann-De Ridder, Aude and Gehrmann, Thomas and Glover, Nigel and Marcoli, Matteo and Preuss, Christian T.",
    title = "{Jet Rates in Higgs Boson Decay at Third Order in QCD}",
    eprint = "2502.17333",
    archivePrefix = "arXiv",
    primaryClass = "hep-ph",
    reportNumber = "IPPP/25/10, ZU-TH 09/25, MCNET-25-03",
    doi = "10.1103/1znh-sm96",
    journal = "Phys. Rev. Lett.",
    volume = "134",
    pages = "251905",
    year = "2025"
}

@article{Braun-White:2023sgd,
    author = "Braun-White, Oscar and Glover, Nigel and Preuss, Christian T.",
    title = "{A general algorithm to build real-radiation antenna functions for higher-order calculations}",
    eprint = "2302.12787",
    archivePrefix = "arXiv",
    primaryClass = "hep-ph",
    reportNumber = "IPPP 23/7",
    doi = "10.1007/JHEP06(2023)065",
    journal = "JHEP",
    volume = "06",
    pages = "065",
    year = "2023"
}

@article{Braun-White:2023zwd,
    author = "Braun-White, Oscar and Glover, Nigel and Preuss, Christian T.",
    title = "{A general algorithm to build mixed real and virtual antenna functions for higher-order calculations}",
    eprint = "2307.14999",
    archivePrefix = "arXiv",
    primaryClass = "hep-ph",
    reportNumber = "IPPP/23/31, ZU-TH 30/23",
    doi = "10.1007/JHEP11(2023)179",
    journal = "JHEP",
    volume = "11",
    pages = "179",
    year = "2023"
}

@article{Fox:2024bfp,
    author = "Fox, Elliot and Glover, Nigel and Marcoli, Matteo",
    title = "{Generalised antenna functions for higher-order calculations}",
    eprint = "2410.12904",
    archivePrefix = "arXiv",
    primaryClass = "hep-ph",
    reportNumber = "IPPP/24/63, IPPP/24/63",
    doi = "10.1007/JHEP12(2024)225",
    journal = "JHEP",
    volume = "12",
    pages = "225",
    year = "2024"
}

@article{NNLOJET:2025rno,
    author = "Huss, A. and others",
    collaboration = "NNLOJET",
    title = "{NNLOJET: a parton-level event generator for jet cross sections at NNLO QCD accuracy}",
    eprint = "2503.22804",
    archivePrefix = "arXiv",
    primaryClass = "hep-ph",
    reportNumber = "CERN-TH-2025-012, IPPP/25/09, ZU-TH 11/25",
    month = "3",
    year = "2025"
}

@article{Gehrmann-DeRidder:2023uld,
    author = "Gehrmann-De Ridder, Aude and Preuss, Christian T. and Williams, Ciaran",
    title = "{Four-jet event shapes in hadronic Higgs decays}",
    eprint = "2310.09354",
    archivePrefix = "arXiv",
    primaryClass = "hep-ph",
    reportNumber = "ZU-TH 59/23",
    doi = "10.1007/JHEP03(2024)104",
    journal = "JHEP",
    volume = "03",
    pages = "104",
    year = "2024"
}

@article{Currie:2013vh,
    author = "Currie, James and Glover, E. W. N. and Wells, Steven",
    title = "{Infrared Structure at NNLO Using Antenna Subtraction}",
    eprint = "1301.4693",
    archivePrefix = "arXiv",
    primaryClass = "hep-ph",
    reportNumber = "IPPP-12-82, ZU-TH-26-12",
    doi = "10.1007/JHEP04(2013)066",
    journal = "JHEP",
    volume = "04",
    pages = "066",
    year = "2013"
}

@article{Weinzierl:2009nz,
    author = "Weinzierl, Stefan",
    title = "{The infrared structure of $e^+ e^- \to 3$ jets at NNLO reloaded}",
    eprint = "0904.1145",
    archivePrefix = "arXiv",
    primaryClass = "hep-ph",
    reportNumber = "MZ-TH-09-12",
    doi = "10.1088/1126-6708/2009/07/009",
    journal = "JHEP",
    volume = "07",
    pages = "009",
    year = "2009"
}

@article{Hoang:2014wka,
    author = "Hoang, Andr{\'e} H. and Kolodrubetz, Daniel W. and Mateu, Vicent and Stewart, Iain W.",
    title = "{$C$-parameter distribution at N$^3$LL' including power corrections}",
    eprint = "1411.6633",
    archivePrefix = "arXiv",
    primaryClass = "hep-ph",
    reportNumber = "UWTHPH-2014-07, MIT-CTP-4596, LPN14-123",
    doi = "10.1103/PhysRevD.91.094017",
    journal = "Phys. Rev. D",
    volume = "91",
    pages = "094017",
    year = "2015"
}

@article{Banfi:2018mcq,
    author = "Banfi, Andrea and El-Menoufi, Basem Kamal and Monni, Pier Francesco",
    title = "{The Sudakov radiator for jet observables and the soft physical coupling}",
    eprint = "1807.11487",
    archivePrefix = "arXiv",
    primaryClass = "hep-ph",
    doi = "10.1007/JHEP01(2019)083",
    journal = "JHEP",
    volume = "01",
    pages = "083",
    year = "2019"
}

@article{Bhattacharya:2022dtm,
    author = "Bhattacharya, Arindam and Schwartz, Matthew D. and Zhang, Xiaoyuan",
    title = "{Sudakov shoulder resummation for thrust and heavy jet mass}",
    eprint = "2205.05702",
    archivePrefix = "arXiv",
    primaryClass = "hep-ph",
    doi = "10.1103/PhysRevD.106.074011",
    journal = "Phys. Rev. D",
    volume = "106",
    pages = "074011",
    year = "2022"
}

@article{Luisoni:2020efy,
    author = "Luisoni, Gionata and Monni, Pier Francesco and Salam, Gavin P.",
    title = "{$C$-parameter hadronisation in the symmetric 3-jet limit and impact on $\alpha_s$ fits}",
    eprint = "2012.00622",
    archivePrefix = "arXiv",
    primaryClass = "hep-ph",
    reportNumber = "CERN-TH-2019-155",
    doi = "10.1140/epjc/s10052-021-08941-z",
    journal = "Eur. Phys. J. C",
    volume = "81",
    pages = "158",
    year = "2021"
}

@article{Caola:2021kzt,
    author = "Caola, Fabrizio and Ferrario Ravasio, Silvia and Limatola, Giovanni and Melnikov, Kirill and Nason, Paolo",
    title = "{On linear power corrections in certain collider observables}",
    eprint = "2108.08897",
    archivePrefix = "arXiv",
    primaryClass = "hep-ph",
    reportNumber = "OUTP-21-21P, TTP21-026, P3H-21-056",
    doi = "10.1007/JHEP01(2022)093",
    journal = "JHEP",
    volume = "01",
    pages = "093",
    year = "2022"
}

@article{Caola:2022vea,
    author = "Caola, Fabrizio and Ferrario Ravasio, Silvia and Limatola, Giovanni and Melnikov, Kirill and Nason, Paolo and Ozcelik, Melih Arslan",
    title = "{Linear power corrections to e$^{+}$e$^{-}$ shape variables in the three-jet region}",
    eprint = "2204.02247",
    archivePrefix = "arXiv",
    primaryClass = "hep-ph",
    reportNumber = "OUTP-22-04P, TTP22-022, P3H-22-036, MPP-2022-36",
    doi = "10.1007/JHEP12(2022)062",
    journal = "JHEP",
    volume = "12",
    pages = "062",
    year = "2022"
}

@article{Agarwal:2023fdk,
    author = "Agarwal, Neelima and van Beekveld, Melissa and Laenen, Eric and Mishra, Shubham and Mukhopadhyay, Ayan and Tripathi, Anurag",
    title = "{Next-to-leading power corrections to event-shape variables}",
    eprint = "2306.17601",
    archivePrefix = "arXiv",
    primaryClass = "hep-ph",
    doi = "10.1007/s12043-024-02743-0",
    journal = "Pramana",
    volume = "98",
    pages = "60",
    year = "2024"
}

@article{Coloretti:2022jcl,
    author = "Coloretti, Guglielmo and Gehrmann-De Ridder, Aude and Preuss, Christian T.",
    title = "{QCD predictions for event-shape distributions in hadronic Higgs decays}",
    eprint = "2202.07333",
    archivePrefix = "arXiv",
    primaryClass = "hep-ph",
    reportNumber = "ZU-TH 05/22",
    doi = "10.1007/JHEP06(2022)009",
    journal = "JHEP",
    volume = "06",
    pages = "009",
    year = "2022"
}

@article{DelDuca:2004wt,
    author = "Del Duca, Vittorio and Frizzo, Alberto and Maltoni, Fabio",
    title = "{Higgs boson production in association with three jets}",
    eprint = "hep-ph/0404013",
    archivePrefix = "arXiv",
    reportNumber = "DFTT-11-2004",
    doi = "10.1088/1126-6708/2004/05/064",
    journal = "JHEP",
    volume = "05",
    pages = "064",
    year = "2004"
}

@article{Dixon:2009uk,
    author = "Dixon, Lance J. and Sofianatos, Yorgos",
    title = "{Analytic one-loop amplitudes for a Higgs boson plus four partons}",
    eprint = "0906.0008",
    archivePrefix = "arXiv",
    primaryClass = "hep-ph",
    reportNumber = "SLAC-PUB-13652",
    doi = "10.1088/1126-6708/2009/08/058",
    journal = "JHEP",
    volume = "08",
    pages = "058",
    year = "2009"
}

@article{Badger:2009hw,
    author = "Badger, Simon and Nigel Glover, E. W. and Mastrolia, Pierpaolo and Williams, Ciaran",
    title = "{One-loop Higgs plus four gluon amplitudes: Full analytic results}",
    eprint = "0909.4475",
    archivePrefix = "arXiv",
    primaryClass = "hep-ph",
    reportNumber = "IPPP-09-58, CERN-PH-TH-2009-163, DESY-09-138",
    doi = "10.1007/JHEP01(2010)036",
    journal = "JHEP",
    volume = "01",
    pages = "036",
    year = "2010"
}

@article{Badger:2009vh,
    author = "Badger, Simon and Campbell, John M. and Ellis, R. Keith and Williams, Ciaran",
    title = "{Analytic results for the one-loop NMHV Hqqgg amplitude}",
    eprint = "0910.4481",
    archivePrefix = "arXiv",
    primaryClass = "hep-ph",
    reportNumber = "DESY-09-180, FERMILAB-PUB-09-505-T, IPPP-09-86",
    doi = "10.1088/1126-6708/2009/12/035",
    journal = "JHEP",
    volume = "12",
    pages = "035",
    year = "2009"
}

@article{Chen:2014gva,
    author = "Chen, X. and Gehrmann, T. and Glover, E. W. N. and Jaquier, M.",
    title = "{Precise QCD predictions for the production of Higgs + jet final states}",
    eprint = "1408.5325",
    archivePrefix = "arXiv",
    primaryClass = "hep-ph",
    reportNumber = "IPPP-14-64, ZU-TH-27-14",
    doi = "10.1016/j.physletb.2014.11.021",
    journal = "Phys. Lett. B",
    volume = "740",
    pages = "147--150",
    year = "2015"
}

@article{Ahmed:2014pka,
    author = "Ahmed, Taushif and Mahakhud, Maguni and Mathews, Prakash and Rana, Narayan and Ravindran, V.",
    title = "{Two-loop QCD corrections to Higgs $\to b+\overline{b}+g$ amplitude}",
    eprint = "1405.2324",
    archivePrefix = "arXiv",
    primaryClass = "hep-ph",
    reportNumber = "HRI-RECAPP-2014-010",
    doi = "10.1007/JHEP08(2014)075",
    journal = "JHEP",
    volume = "08",
    pages = "075",
    year = "2014"
}

@article{Jones:2003yv,
    author = "Jones, Roger W. L. and Ford, Matthew and Salam, Gavin P. and Stenzel, Hasko and Wicke, Daniel",
    title = "{Theoretical uncertainties on $\alpha_s$ from event shape variables in $e^+ e^-$ annihilations}",
    eprint = "hep-ph/0312016",
    archivePrefix = "arXiv",
    doi = "10.1088/1126-6708/2003/12/007",
    journal = "JHEP",
    volume = "12",
    pages = "007",
    year = "2003"
}

@article{Wilczek:1977zn,
    author = "Wilczek, Frank",
    title = "{Decays of Heavy Vector Mesons Into Higgs Particles}",
    reportNumber = "FERMILAB-PUB-77-080-T",
    doi = "10.1103/PhysRevLett.39.1304",
    journal = "Phys. Rev. Lett.",
    volume = "39",
    pages = "1304",
    year = "1977"
}

@article{Catani:1997xc,
    author = "Catani, S. and Webber, B. R.",
    title = "{Infrared safe but infinite: Soft gluon divergences inside the physical region}",
    eprint = "hep-ph/9710333",
    archivePrefix = "arXiv",
    reportNumber = "CAVENDISH-HEP-97-10, LPTHE-ORSAY-97-39",
    doi = "10.1088/1126-6708/1997/10/005",
    journal = "JHEP",
    volume = "10",
    pages = "005",
    year = "1997"
}

@article{Knobbe:2023njd,
    author = "Knobbe, Max and Krauss, Frank and Reichelt, Daniel and Schumann, Steffen",
    title = "{Measuring hadronic Higgs boson branching ratios at future lepton colliders}",
    eprint = "2306.03682",
    archivePrefix = "arXiv",
    primaryClass = "hep-ph",
    reportNumber = "IPPP/23/26, MCNET-23-06",
    doi = "10.1140/epjc/s10052-024-12430-4",
    journal = "Eur. Phys. J. C",
    volume = "84",
    pages = "83",
    year = "2024"
}

@article{Banfi:2004yd,
    author = "Banfi, Andrea and Salam, Gavin P. and Zanderighi, Giulia",
    title = "{Principles of general final-state resummation and automated implementation}",
    eprint = "hep-ph/0407286",
    archivePrefix = "arXiv",
    reportNumber = "FERMILAB-PUB-04-116-T, LPTHE-04-16, NIKHEF-2004-005",
    doi = "10.1088/1126-6708/2005/03/073",
    journal = "JHEP",
    volume = "03",
    pages = "073",
    year = "2005"
}

@article{Arpino:2019ozn,
    author = "Arpino, Luke and Banfi, Andrea and El-Menoufi, Basem Kamal",
    title = "{Near-to-planar three-jet events at NNLL accuracy}",
    eprint = "1912.09341",
    archivePrefix = "arXiv",
    primaryClass = "hep-ph",
    doi = "10.1007/JHEP07(2020)171",
    journal = "JHEP",
    volume = "07",
    pages = "171",
    year = "2020"
}

@article{Fox:2025qmp,
    author = "Fox, Elliot and Gehrmann-De Ridder, Aude and Gehrmann, Thomas and Glover, Nigel and Marcoli, Matteo and Preuss, Christian T.",
    title = "{Precise predictions for event shapes in hadronic Higgs decays}",
    eprint = "2508.14282",
    archivePrefix = "arXiv",
    primaryClass = "hep-ph",
    reportNumber = "IPPP/25/53, ZU-TH 52/25, MCNET-25-17",
    doi = "10.1007/JHEP11(2025)168",
    journal = "JHEP",
    volume = "11",
    pages = "168",
    year = "2025"
}

@article{Ju:2023dfa,
    author = "Ju, Wan-Li and Xu, Yongqi and Yang, Li Lin and Zhou, Bin",
    title = "{Thrust distribution in Higgs decays up to the fifth logarithmic order}",
    eprint = "2301.04294",
    archivePrefix = "arXiv",
    primaryClass = "hep-ph",
    doi = "10.1103/PhysRevD.107.114034",
    journal = "Phys. Rev. D",
    volume = "107",
    pages = "114034",
    year = "2023"
}

@article{Gehrmann-DeRidder:2024avt,
    author = "Gehrmann-De Ridder, Aude and Preuss, Christian T. and Reichelt, Daniel and Schumann, Steffen",
    title = "{NLO+NLL' accurate predictions for three-jet event shapes in hadronic Higgs decays}",
    eprint = "2403.06929",
    archivePrefix = "arXiv",
    primaryClass = "hep-ph",
    reportNumber = "MCNET-24-03, ZU-TH 16/24, IPPP/24/09",
    doi = "10.1007/JHEP07(2024)160",
    journal = "JHEP",
    volume = "07",
    pages = "160",
    year = "2024"
}

@article{Ellis:1980wv,
    author = "Ellis, R. Keith and Ross, D. A. and Terrano, A. E.",
    title = "{The Perturbative Calculation of Jet Structure in $e^+ e^-$ Annihilation}",
    reportNumber = "CALT-68-785",
    doi = "10.1016/0550-3213(81)90165-6",
    journal = "Nucl. Phys. B",
    volume = "178",
    pages = "421--456",
    year = "1981"
}

@article{Nason:2023asn,
    author = "Nason, Paolo and Zanderighi, Giulia",
    title = "{Fits of {\ensuremath{\alpha}}$_{s}$ using power corrections in the three-jet region}",
    eprint = "2301.03607",
    archivePrefix = "arXiv",
    primaryClass = "hep-ph",
    doi = "10.1007/JHEP06(2023)058",
    journal = "JHEP",
    volume = "06",
    pages = "058",
    year = "2023"
}

@article{Nason:2025qbx,
    author = "Nason, Paolo and Zanderighi, Giulia",
    title = "{Fits of {\ensuremath{\alpha}}$_{s}$ from event-shapes in the three-jet region: extension to all energies}",
    eprint = "2501.18173",
    archivePrefix = "arXiv",
    primaryClass = "hep-ph",
    reportNumber = "MPP-2025-9",
    doi = "10.1007/JHEP06(2025)200",
    journal = "JHEP",
    volume = "06",
    pages = "200",
    year = "2025"
}

@article{FCC:2018byv,
    author = "Abada, A. and others",
    collaboration = "FCC",
    title = "{FCC Physics Opportunities}: {Future Circular Collider Conceptual Design Report Volume 1}",
    reportNumber = "CERN-ACC-2018-0056",
    doi = "10.1140/epjc/s10052-019-6904-3",
    journal = "Eur. Phys. J. C",
    volume = "79",
    pages = "474",
    year = "2019"
}

@article{FCC:2018evy,
    author = "Abada, A. and others",
    collaboration = "FCC",
    title = "{FCC-ee: The Lepton Collider}: {Future Circular Collider Conceptual Design Report Volume 2}",
    reportNumber = "CERN-ACC-2018-0057",
    doi = "10.1140/epjst/e2019-900045-4",
    journal = "Eur. Phys. J. ST",
    volume = "228",
    pages = "261--623",
    year = "2019"
}

@article{CEPCStudyGroup:2018ghi,
    author = "Dong, Mingyi and others",
    editor = "Guimar\~aes da Costa, Jo\~ao Barreiro and others",
    collaboration = "CEPC Study Group",
    title = "{CEPC Conceptual Design Report: Volume 2 - Physics \& Detector}",
    eprint = "1811.10545",
    archivePrefix = "arXiv",
    primaryClass = "hep-ex",
    reportNumber = "IHEP-CEPC-DR-2018-02, IHEP-EP-2018-01, IHEP-TH-2018-01",
    month = "11",
    year = "2018"
}

@article{ILC:2013jhg,
    author = "Baer, Howard and others",
    collaboration = "ILC",
    title = "{The International Linear Collider Technical Design Report - Volume 2: Physics}",
    eprint = "1306.6352",
    archivePrefix = "arXiv",
    primaryClass = "hep-ph",
    reportNumber = "ILC-REPORT-2013-040, ANL-HEP-TR-13-20, BNL-100603-2013-IR, IRFU-13-59, CERN-ATS-2013-037, COCKCROFT-13-10, CLNS-13-2085, DESY-13-062, FERMILAB-TM-2554, IHEP-AC-ILC-2013-001, INFN-13-04-LNF, JAI-2013-001, JINR-E9-2013-35, JLAB-R-2013-01, KEK-REPORT-2013-1, KNU-CHEP-ILC-2013-1, LLNL-TR-635539, SLAC-R-1004, ILC-HIGRADE-REPORT-2013-003",
    month = "6",
    year = "2013"
}

@article{Hoang:2025uaa,
    author = "Hoang, Andre H. and Mateu, Vicent and Schwartz, Matthew D. and Stewart, Iain W.",
    title = "{Precision e$^{+}$e$^{-}$ hemisphere masses in the dijet region with power corrections}",
    eprint = "2506.09130",
    archivePrefix = "arXiv",
    primaryClass = "hep-ph",
    reportNumber = "UWThPh-2025-5, MIT-CTP 5028",
    doi = "10.1007/JHEP09(2025)092",
    journal = "JHEP",
    volume = "09",
    pages = "092",
    year = "2025"
}

@article{JADE:1999zar,
    author = "Pfeifenschneider, P. and others",
    collaboration = "JADE, OPAL",
    title = "{QCD analyses and determinations of alpha(s) in e+ e- annihilation at energies between 35-GeV and 189-GeV}",
    eprint = "hep-ex/0001055",
    archivePrefix = "arXiv",
    reportNumber = "CERN-EP-99-175",
    doi = "10.1007/s100520000432",
    journal = "Eur. Phys. J. C",
    volume = "17",
    pages = "19--51",
    year = "2000"
}

@article{L3:2004cdh,
    author = "Achard, P. and others",
    collaboration = "L3",
    title = "{Studies of hadronic event structure in $e^{+} e^{-}$ annihilation from 30-GeV to 209-GeV with the L3 detector}",
    eprint = "hep-ex/0406049",
    archivePrefix = "arXiv",
    reportNumber = "CERN-PH-EP-2004-024, CERN-EP-PH-2004-024",
    doi = "10.1016/j.physrep.2004.07.002",
    journal = "Phys. Rept.",
    volume = "399",
    pages = "71--174",
    year = "2004"
}

@article{Yan:2023xsd,
    author = "Yan, Bin and Lee, Christopher",
    title = "{Probing light quark Yukawa couplings through angularity distributions in Higgs boson decay}",
    eprint = "2311.12556",
    archivePrefix = "arXiv",
    primaryClass = "hep-ph",
    reportNumber = "LA-UR-23-32929",
    doi = "10.1007/JHEP03(2024)123",
    journal = "JHEP",
    volume = "03",
    pages = "123",
    year = "2024"
}

@article{vanBeekveld:2024wws,
    author = "van Beekveld, Melissa and others",
    title = "{New Standard for the Logarithmic Accuracy of Parton Showers}",
    eprint = "2406.02661",
    archivePrefix = "arXiv",
    primaryClass = "hep-ph",
    reportNumber = "CERN-TH-2024-057, OUTP-24-03P",
    doi = "10.1103/PhysRevLett.134.011901",
    journal = "Phys. Rev. Lett.",
    volume = "134",
    number = "1",
    pages = "011901",
    year = "2025"
}

@article{Fox:2025txz,
	author = "Fox, Elliot and Gehrmann-De Ridder, Aude and Gehrmann, Thomas and Glover, Nigel and Marcoli, Matteo and Preuss, Christian T.",
	title = "{The Thrust Distribution at NNLO+NNLL in Higgs Decays to Quarks and Gluons}",
	eprint = "2510.11665",
	archivePrefix = "arXiv",
	primaryClass = "hep-ph",
	reportNumber = "IPPP/25/64, ZU-TH 67/25, MCNET-25-25",
	month = "10",
	year = "2025"
}

@article{Gardi:2003iv,
    author = "Gardi, Einan and Magnea, Lorenzo",
    title = "{The C parameter distribution in e+ e- annihilation}",
    eprint = "hep-ph/0306094",
    archivePrefix = "arXiv",
    reportNumber = "CERN-TH-2003-129, DFTT-6-2003",
    doi = "10.1088/1126-6708/2003/08/030",
    journal = "JHEP",
    volume = "08",
    pages = "030",
    year = "2003"
}

@article{Hoang:2015hka,
    author = "Hoang, Andr{\'e} H. and Kolodrubetz, Daniel W. and Mateu, Vicent and Stewart, Iain W.",
    title = "{Precise determination of $\alpha_s$ from the $C$-parameter distribution}",
    eprint = "1501.04111",
    archivePrefix = "arXiv",
    primaryClass = "hep-ph",
    reportNumber = "UWTHPH-2015-1, MIT-CTP-4630, LPN14-128",
    doi = "10.1103/PhysRevD.91.094018",
    journal = "Phys. Rev. D",
    volume = "91",
    number = "9",
    pages = "094018",
    year = "2015"
}

@article{Bhattacharya:2023qet,
    author = "Bhattacharya, Arindam and Michel, Johannes K. L. and Schwartz, Matthew D. and Stewart, Iain W. and Zhang, Xiaoyuan",
    title = "{NNLL resummation of Sudakov shoulder logarithms in the heavy jet mass distribution}",
    eprint = "2306.08033",
    archivePrefix = "arXiv",
    primaryClass = "hep-ph",
    reportNumber = "MIT-CTP 5570",
    doi = "10.1007/JHEP11(2023)080",
    journal = "JHEP",
    volume = "11",
    pages = "080",
    year = "2023"
}

@article{Chien:2010kc,
    author = "Chien, Yang-Ting and Schwartz, Matthew D.",
    title = "{Resummation of heavy jet mass and comparison to LEP data}",
    eprint = "1005.1644",
    archivePrefix = "arXiv",
    primaryClass = "hep-ph",
    doi = "10.1007/JHEP08(2010)058",
    journal = "JHEP",
    volume = "08",
    pages = "058",
    year = "2010"
}

@article{Buonocore:2023mne,
    author = "Buonocore, Luca and Grazzini, Massimiliano and Guadagni, Flavio and Rottoli, Luca",
    title = "{Subleading power corrections for event shape variables in $e^+ e^-$ annihilation}",
    eprint = "2311.12768",
    archivePrefix = "arXiv",
    primaryClass = "hep-ph",
    reportNumber = "CERN-TH-2023-209",
    doi = "10.1140/epjc/s10052-024-12788-5",
    journal = "Eur. Phys. J. C",
    volume = "84",
    number = "4",
    pages = "437",
    year = "2024"
}

@article{DelDuca:2016ily,
    author = "Del Duca, Vittorio and Duhr, Claude and Kardos, Adam and Somogyi, G{\'a}bor and Sz{\H{o}}r, Zolt{\'a}n and Tr{\'o}cs{\'a}nyi, Zolt{\'a}n and Tulip{\'a}nt, Zolt{\'a}n",
    title = "{Jet production in the CoLoRFulNNLO method: event shapes in electron-positron collisions}",
    eprint = "1606.03453",
    archivePrefix = "arXiv",
    primaryClass = "hep-ph",
    reportNumber = "CERN-TH-2016-138, CP3-16-29, NSF-KITP-16-084",
    doi = "10.1103/PhysRevD.94.074019",
    journal = "Phys. Rev. D",
    volume = "94",
    number = "7",
    pages = "074019",
    year = "2016"
}

@article{Zhu:2023oka,
    author = "Zhu, Jiawei and Song, Yujin and Gao, Jun and Kang, Daekyoung and Maji, Tanmay",
    title = "{Angularity in Higgs boson decays via H{\textrightarrow}gg at NNLL' accuracy*}",
    eprint = "2311.07282",
    archivePrefix = "arXiv",
    primaryClass = "hep-ph",
    doi = "10.1088/1674-1137/ad94e0",
    journal = "Chin. Phys. C",
    volume = "49",
    number = "2",
    pages = "023106",
    year = "2025"
}

@article{Caletti:2025pot,
    author = "Caletti, Simone and Gehrmann-De Ridder, Aude and Marcoli, Matteo",
    title = "{NNLO QCD predictions for jet observables in ZH production at electron-positron colliders}",
    eprint = "2510.20485",
    archivePrefix = "arXiv",
    primaryClass = "hep-ph",
    reportNumber = "IPPP/25/70, ZU-TH 68/25",
    month = "10",
    year = "2025"
}

@article{Gardi:2002bg,
    author = "Gardi, Einan and Rathsman, Johan",
    title = "{The Thrust and heavy jet mass distributions in the two jet region}",
    eprint = "hep-ph/0201019",
    archivePrefix = "arXiv",
    reportNumber = "CERN-TH-2001-371, TSL-ISV-2001-0258",
    doi = "10.1016/S0550-3213(02)00502-3",
    journal = "Nucl. Phys. B",
    volume = "638",
    pages = "243--287",
    year = "2002"
}

@article{Dokshitzer:1997ew,
    author = "Dokshitzer, Yuri L. and Webber, B. R.",
    title = "{Power corrections to event shape distributions}",
    eprint = "hep-ph/9704298",
    archivePrefix = "arXiv",
    reportNumber = "CAVENDISH-HEP-97-2",
    doi = "10.1016/S0370-2693(97)00573-X",
    journal = "Phys. Lett. B",
    volume = "404",
    pages = "321--327",
    year = "1997"
}

@article{Benitez:2025vsp,
    author = "Benitez, Miguel A. and Bhattacharya, Arindam and Hoang, Andre H. and Mateu, Vicent and Schwartz, Matthew D. and Stewart, Iain W. and Zhang, Xiaoyuan",
    title = "{A Precise Determination of $\alpha_s$ from the Heavy Jet Mass Distribution}",
    eprint = "2502.12253",
    archivePrefix = "arXiv",
    primaryClass = "hep-ph",
    reportNumber = "MIT-CTP 5840, UWTHPH 2025-7",
    month = "2",
    year = "2025"
}

\end{document}